\documentclass[structabstract]{aa}  
\usepackage{graphicx}
\usepackage{txfonts}
\begin{document}

   \title{The intracluster magnetic field power spectrum in A2199}

   \subtitle{}
\author{  V. Vacca\inst{1,2}
          \and
          M. Murgia\inst{2}
          \and
          F. Govoni\inst{2}
          \and
          L. Feretti\inst{3}
         \and
          G. Giovannini\inst{3,4}
         \and
          R. A. Perley\inst{5}
         \and 
         G. B. Taylor\inst{5,6}
                   }
\institute{
              Dipartimento di Fisica,
           Universit\`a degli studi di Cagliari, Cittadella Universitaria, 
           I--09042 Monserrato (CA), Italy
           \and
              INAF - Osservatorio Astronomico di Cagliari,
              Poggio dei Pini, Strada 54, I--09012 Capoterra (CA), Italy 
           \and   
              INAF - Istituto di Radioastronomia, 
              Via Gobetti 101, I--40129 Bologna, Italy
           \and 
              Dipartimento di Astronomia, 
              Univ. Bologna, Via Ranzani 1, I--40127 Bologna, Italy
          \and
              Adjunct Astronomer at the National Radio Astronomy Observatory, 
              Socorro, NM 87801 USA
          \and
              Department of Physics and Astronomy, 
              University of New Mexico, Albuquerque NM, 87131, USA
                }


 %
  \abstract
  {}
   {We investigate the magnetic field power spectrum in the cool core 
     galaxy cluster A2199 by analyzing the polarized emission of the central radio 
     source 3C\,338.}
   {The polarized radiation from the radio emitting plasma is modified by the Faraday rotation 
     as it passes through the magneto-ionic intracluster medium.
     We use Very Large Array observations between 1665 and 8415\,{\rm MHz} to produce detailed 
     Faraday rotation measure and fractional polarization images of the radio galaxy. We simulate 
     Gaussian random three-dimensional magnetic field models with different power-law power spectra 
     and we assume that the field strength decreases radially with the thermal gas density as 
     $n_{\rm e}^{\rm \eta}$. By comparing the synthetic and the observed images with a Bayesian approach, 
     we constrain the strength and structure of the magnetic field associated with the intracluster medium.}
 {We find that the Faraday rotation toward 3C\,338 in A2199 is consistent with a magnetic field power law 
   power spectrum characterized by an index $n$=(2.8$\pm$1.3) between a maximum and a minimum scale of 
   fluctuation of $\Lambda_{\rm max}$=(35$\pm$28)\,{\rm kpc} and $\Lambda_{\rm min}$=(0.7$\pm$0.1)\,{\rm kpc}, 
   respectively. 
      By including in the modeling X-ray cavities coincident with the radio galaxy lobes, we find 
   a magnetic field strength of $\langle B_{\rm 0}\rangle$=(11.7$\pm$9.0)\,$\mu${\rm G} 
   at the cluster center. Further out, the field decreases with the radius following the gas density to the power of $\eta$=(0.9$\pm$0.5). 
 }
  {}
 

   \maketitle

\begin{table*}
\caption{Details of the VLA observations of 3C\,338.}             
\label{table1}      
\centering          
\begin{tabular}{c c c c c c}     
\hline\hline       
Obs. frequency, $\nu$               & Bandwidth    &     VLA Configuration  &   Time &   Data        & Project\\
({\rm MHz})               &  ({\rm MHz})       &        ~     &    (h)   &     ~         &   ~    \\
\hline
1665                &     50       &        A     &     8.5  &   91-Jun-18/19&  GG0005\\
4585/4685/4885      &     50       &        B     &     7.0  &   89-Apr-03   &  AG0269\\
4885                &     50       &        B     &     1.5   &   87-Dec-21   &  AS0309\\
4985                &     50       &        C     &     3.0  &   94-Nov-17   &  BG0012\\
4985                &     50       &        AB    &     3.5  &   95-Sep-11/12&  BV0017\\
8415                &     50       &        C     &     3.0  &   94-Nov-17   &  BG0012\\
8415                &     50       &        BC    &     3.8  &   00-Feb-26   &  GG0038\\
\hline 
\multicolumn{6}{l}{\scriptsize Col. 1: Observing frequency; Col. 2: Observing bandwidth;}\\
\multicolumn{6}{l}{\scriptsize Col. 3: VLA configuration; Col. 4: Time on source; 
Col. 5: Dates of observations; Col. 6: VLA project name.}\\     
\end{tabular}
\end{table*}

\begin{table*}
\caption{Relevant parameters of the total intensity and polarization images between 1665 and 8415\,{\rm MHz}.}  
\label{table2}      
\centering          
\begin{tabular}{c c c c c c}     
\hline\hline       
$\nu$& Beam           &$\sigma$(I)&$\sigma$(Q)&$\sigma$(U)&   $S_{\rm \nu}$  \\
({\rm MHz})&($\arcsec\times\arcsec$ )     &(mJy/beam) &(mJy/beam) &(mJy/beam) & (mJy)     \\
\hline 
1665 & 2.5$\times$2.5 &  0.25     &  0.029    & 0.029   &1490$\pm$ 40\\
4585 & 2.5$\times$2.5 &  0.06     &  0.033    & 0.032   &520$\pm$ 20\\
4685 & 2.5$\times$2.5 &  0.06     &  0.025    & 0.026   &470$\pm$ 10\\
4885 & 2.5$\times$2.5 &  0.06     &  0.026    & 0.027   &440$\pm$ 10\\
4985 & 2.5$\times$2.5 &  0.08     &  0.037    & 0.037   &400$\pm$ 10\\
8415 & 2.5$\times$2.5 &  0.02     &  0.015    & 0.015   &170$\pm$ 10\\
\hline    
 \multicolumn{6}{l}{\scriptsize Col. 1: Observation frequency; Col. 2: FWHM; Col. 3,4,5: RMS noise of the 
   I, Q, U images;}\\
\multicolumn{6}{l}{\scriptsize Col. 6: Flux density.}\\         
\end{tabular}
\end{table*}

\section{Introduction}

In the past decade, significant progress has been made in 
characterizing the properties of the intracluster magnetic field.
Most of what we know about intracluster magnetic field strength and
structure is derived from the study of diffuse synchrotron emission
(radio halos, mini-halos, and relics) and Faraday rotation measures of
polarized radio galaxies located inside or behind galaxy clusters
(e.g., Carilli \& Taylor 2002; Govoni \& Feretti 2004; Ferrari et
al. 2008, and references therein).

The study of the Faraday rotation of cluster radio galaxies provides a
detailed view of the intracluster magnetic field on scales
$\lesssim$100\,{\rm kpc}.  Linearly polarized radiation propagating through
a magnetized plasma experiences a rotation of the plane of
polarization that is proportional to the thermal gas density and the
magnetic field strength along the line-of-sight (e.g., Burn 1966;
Taylor et al. 1999). Indeed, it is possible to obtain important
information about the intracluster magnetic fields by combining
polarization images of radio sources located inside or behind galaxy
clusters with X-ray observations of the thermal gas.

The rotation measure (RM) distributions seen over extended radio
sources are generally patchy, indicating that the intracluster magnetic
fields are not regularly ordered, but instead have turbulent
structures on linear scales as small as a few {\rm kpc} or less, both in merging
and in relaxed clusters.  Actually, magnetic field fluctuations seem
to cover a wide range of spatial scales. In a few galaxy clusters
containing radio sources with very detailed rotation measure images,
the magnetic field power spectrum has been estimated (En{\ss}lin \&
Vogt 2003; Vogt \& En{\ss}lin 2003; Murgia et al. 2004; Vogt \&
En{\ss}lin 2005; Govoni et al. 2006; Guidetti et al. 2008; Laing et
al. 2008; Bonafede et al. 2010; Guidetti et al. 2010; Kuchar 
\& En{\ss}lin 2011).  The rotation
measure studies generally show central magnetic field strengths of a
few $\mu${\rm G} in merging galaxy clusters. Higher values of about
10-40\,$\mu${\rm G} are typical of relaxed cooling core clusters (e.g.
Dreher et al. 1987; Allen et al. 2001; Taylor et al. 2002), where the
observed extreme RM magnitudes appear to be roughly proportional to
the cooling flow rate (Taylor et al. 2002).  In this paper, we
investigate the magnetic field power spectrum in the nearby galaxy
cluster A2199 by analyzing multifrequency polarization observations
of the central radio source 3C\,338 taken with the Very Large Array
(VLA).

\emph{Chandra} observations by Johnstone et al. (2002) revealed X-ray
cavities in the cluster center associated with the radio lobes. The
temperature of the intracluster medium decreases from 4.2\,keV to
1.6\,keV over radii from 100\,{\rm kpc} to 5\,{\rm kpc}, implying a drop in the
radiative cooling time from 7\,Gyr to 0.1\,Gyr. These features seem to
be consistent with a cooling flow, even if the action of radiative
cooling should cool the central ICM to \emph{kT}$<$1\,keV. Therefore,
as in many other cool core clusters, in A2199, it has been
proposed that some heating mechanism (e.g., Tucker \& Rosner 1983;
Gaetz 1989; David et al. 2000) prevents the ICM from cooling down to
temperatures below 1\,keV. Kawano et al. (2003) suggest that previous
AGN activity could be responsible for such heating in A2199.

The radio source 3C\,338 (otherwise known as B2 1626+39) is associated
with the multiple nuclei cD galaxy NGC 6166, the brightest galaxy at
the center of the cluster (Burns et al. 1983, Fanti et al. 1986), and
it is classified as a restarting Fanaroff-Riley type I radio galaxy
(see Murgia et al. 2011 and references therein).  On parsec scales,
3C\,338 was the first established example of a symmetric, two-sided 
source in a
radio galaxy (Feretti et al. 1993), consisting of a compact core and
two symmetric relativistic jets (Giovannini et al. 2001).  On kiloparsec
scales, Burns et al. (1983) and Giovannini et al. (1998) showed that
the radio structure consists of an active region, which includes the
core and two symmetric jets terminating in two faint hot spots and 
two steep-spectrum radio lobes.  The radio lobes are clearly
associated with cavities in the central X-ray emission, as shown by
Johnstone et al. (2002) and Gentile et al. (2007). Burns et al. (1983)
pointed out a displacement between the large-scale structure with
respect to the restarted symmetric jets, which could indicate a
possible motion of the central AGN inside the galaxy, but we speculate
that the displacement could also be due to an interaction between the
old radio lobes with bulk motions in the surrounding medium 
caused by the sloshing of the cluster core (Markevitch
\& Vikhlinin 2007).
 
A2199 is an interesting target for Faraday rotation studies because
the presence of X-ray cavities associated with the radio galaxy lobes
indicates that the rotation of the polarization plane is likely to
occur entirely in the intracluster medium, since comparatively little
thermal gas should be present inside the radio-emitting plasma.  A
previous rotation measure study has been done by Ge \& Owen (1994)
based on 5000\,{\rm MHz} VLA data. These authors show that 3C\,338 radio
emission suffers a significant depolarization and is characterized
by high RM values. By combining the information from this rotation
measure image with deprojected ROSAT data, and assuming a very simple
magnetic field model, Eilek \& Owen (2002) infer an averaged
magnetic field value along the line of sight of 15\,$\mu${\rm G}.
  
In this paper we try to improve upon the previous estimate by analyzing
additional data and by performing a numerical modeling of the intracluster
magnetic field fluctuations. We use VLA observations between 1665 and 8415\,{\rm MHz}
to produce detailed Faraday rotation measure and fractional
polarization images of the radio galaxy. Following Murgia et
al. (2004), we simulate Gaussian random three-dimensional magnetic
field models with different power-law power spectra, and we compare
the synthetic and the observed images in order to constrain the strength and
structure of the magnetic field associated with the intracluster
medium.

In \S\,2 we present the radio observations and the data reduction. In
\S\,3 we describe the polarization properties of 3C\,338. In \S\,4 we
present the Faraday rotation modeling. In sections 5 and 6 we describe
the results of the two-dimensional and three-dimensional
simulations. Finally, in \S\,7 we summarize our conclusions.
Throughout this paper we adopt a $\Lambda$CDM cosmology with $H_{\rm
  0}=71$\,{\rm km s}$^{\rm -1}$\,{\rm Mpc}$^{\rm -1}$, $\Omega_{\rm m}=0.27$, and
$\Omega_{\rm \Lambda}=0.73$. At the distance of 3C\,338 (z=0.0311,
Smith et al. 1997), 1\arcsec\, corresponds to 0.61\,{\rm kpc}.
 
\begin{figure}[ht]
   \centering
   \includegraphics[width=1\columnwidth]{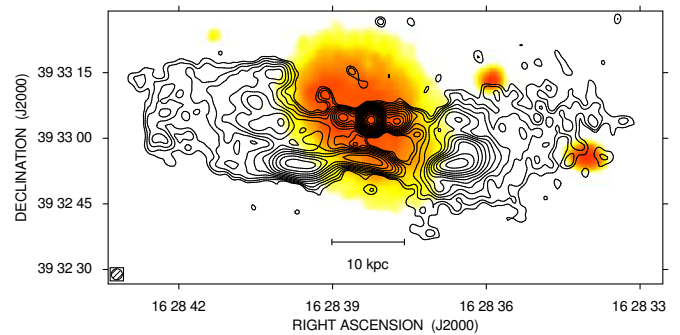}
     \caption{Total intensity radio contours of 3C\,338 at 8415\,{\rm MHz} with an
FWHM of $2.5\arcsec \times 2.5\arcsec$. The first contour level is drawn at 0.06\,mJy/beam
and the rest are spaced by a factor $\sqrt{2}$. The sensitivity (1$\sigma$) is 0.02\,mJy/beam. 
The contours of the radio intensity are overlaid on the DSS2 red plate.}
              \label{figure1}
    \end{figure}

 \begin{figure}[ht]
   \centering
  \includegraphics[width=10cm]{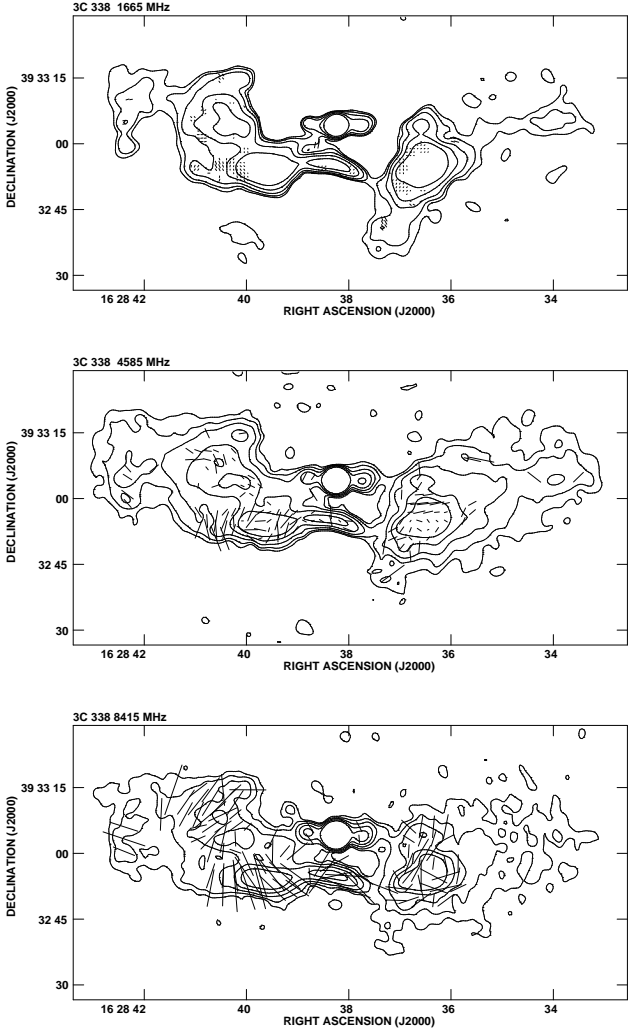}
      \caption{Total intensity contours and polarization vectors at 1665 (\emph{top}), 4585 (\emph{middle}), and 
        8415\,{\rm MHz} (\emph{bottom}). The angular resolution is $2.5\arcsec \times 2.5\arcsec$.  Contour levels 
        start at 3$\sigma_{\rm I}$, and increase by factors of two. The lines give the orientation of the 
        electric vector position angle (E-field) and are proportional in length to the fractional polarization 
        ($1\arcsec \sim$ 5\%). }
              \label{figure2}
    \end{figure}

\section{Radio observations and data reduction}

\label{Radio observations and data reduction}
We present polarimetric archival observations of 3C\,338 at 1665,
4585, 4685, 4885, 4985, and 8415\,{\rm MHz} performed at the VLA in the A,
B, and C configurations between 1987 December and 2000 February. All
observations were made with a bandwidth of 50\,{\rm MHz}. The pointing
center is identified by the coordinates 
RA(J2000)=$16^{\rm h}28^{\rm m}38^{\rm s}$, and 
DEC(J2000)=$+39^{\rm d}33^{\rm m}04^{\rm s}$, and the
details are provided in Table\,\ref{table1}.  

All data were reduced using the NRAO's Astronomical Image Processing
System (AIPS) package.  The radio source 3C\,286 was used both as
primary flux density calibrator and as reference for the absolute
polarization angles. Phase calibration was derived from nearby
sources, periodically observed over a wide range in parallactic angle
to separate the source polarization properties from the antenna polarizations.  
We proceeded with the standard calibration procedures, imaging and self-calibration. Once all the data were calibrated for
each configuration and frequency, all visibilities corresponding to
the same frequency were combined using the task DBCON in order to
improve $(u,v)$ coverage and sensitivity. The combined data were
then self-calibrated to produce the final images.  The images of total
intensity \emph{I} and Stokes parameters \emph{Q} and \emph{U} were
restored with a circular beam of $2.5\arcsec$. The noise levels of
\emph{I}, \emph{U}, and \emph{Q} are summarized in Table
\,\ref{table2}. Polarized intensity (corrected for the positive bias),
$P=\sqrt{Q^{\rm 2}+U^{\rm 2}}$, polarization angle, $\Psi=0.5tan^{\rm -1}(U/Q)$, and
fractional polarization $FPOL=P/I$ images were produced for each
frequency. 

The total intensity contours of the radio galaxy 3C\,338 overlaid on
the DSS2 red plate\footnote{http://archive.eso.org/dss/dss} are showed
in Fig.\,\ref{figure1}. The 8415\,{\rm MHz} radio galaxy emission peaks at
RA(J2000)=$16^{\rm h}28^{\rm m}38^{\rm s}$ and
DEC(J2000)=$+39^{\rm d}33^{\rm m}04^{\rm s}$. It is also the
same location of the X-ray peak to within the image resolution. Therefore
we chose this reference location for our analysis.

\section{Polarization analysis} 
\label{Polarization analysis}
In Fig.\,\ref{figure2} we present the total intensity radio contours
and polarization vectors at 1665, 4585, and 8415\,{\rm MHz} at a resolution
of $2.5\arcsec \times 2.5\arcsec$. The length of the vectors is
proportional to the fractional polarization, while their orientation
is the same as the projected E-field.  The fractional polarization and
the polarization angle were obtained by considering only those pixels
where the fractional polarization is above 3$\sigma_{\rm FPOL}$ at the
corresponding frequency. The total intensity emission of the source
extends out to about 1$\rm \arcmin$ from the cluster center.
The total flux density of the radio source and all the relevant parameters of the $I$, $Q$, and $U$ images are reported in Table\,\ref{table2}. 

\subsection{Rotation measure}
\label{Rotation measure}
\begin{table}
\caption{Statistics of the rotation measure distribution.}             
\label{table3}      
\centering          
\begin{tabular}{c c c c c c}     
\hline\hline  
Component &   $\langle RM\rangle$    & $\sigma_{\rm RM}$ & $\mid RM_{\rm max}\mid$ &  $Err_{\rm fit}$ &$\langle RM^{\rm 2}\rangle^{\rm 1/2}$\\
 ~     & ({\rm rad/m}$^{\rm 2}$) &  ({\rm rad/m}$^{\rm 2}$)  &    ({\rm rad/m}$^{\rm 2}$)     & ({\rm rad/m}$^{\rm 2}$) &  ({\rm rad/m}$^{\rm 2}$)\\
Total  &   -54       &     460       &      903           &     18 &463\\   
E-Lobe  &   66       &     473       &      903           &     20 &478\\   
W-Lobe  &   -251       &     356       &      581           &     16 &436\\ 
\hline    
 \multicolumn{6}{l}{\scriptsize Col.\,1: Component of the source; Col.\,2: Mean value of the RM distribution;}\\
\multicolumn{6}{l}{\scriptsize Col.\,3: Standard deviation of the RM distribution;}\\    
\multicolumn{6}{l}{\scriptsize Col.\,4: Maximum absolute value of the RM distribution;}\\  
\multicolumn{6}{l}{\scriptsize Col.\,5: Mean value of the RM fit error;}\\   
 \multicolumn{6}{l}{\scriptsize Col.\,6: Total power, 
   $\langle RM^{\rm 2}\rangle^{\rm 1/2}=(\langle RM\rangle^{\rm 2}+\sigma_{\rm RM}^2)^{\rm 1/2}$.}
\end{tabular}
\end{table}

The presence of a magnetic field in an ionized plasma creates a
difference in the phase velocities for left versus right circularly
polarized radiation. As a consequence, the polarized emission from a
radio source propagating through the plasma experiences a phase shift
between the two components.  This corresponds to a rotation in the
polarization angle. For a completely foreground screen (Burn 1966), which
we expect here, the rotation is
\begin{equation}
\Psi_{\rm obs}(\lambda)=\Psi_{\rm int}+\lambda ^{\rm 2}\mathrm{\rm RM},
\label{eq1}
\end{equation}
where $\Psi_{\rm obs}(\lambda)$ is the observed position angle at
wavelength $\lambda$, $\Psi_{\rm int}$ the intrinsic polarization
angle of the polarized emission, and RM the rotation measure. Under
this assumption, by observing radio galaxies at different wavelengths,
the rotation measure can be derived from a linear fit of the
polarization position angle $\Psi_{\rm obs}(\lambda)$ versus
$\lambda^{\rm 2}$.  By considering an electron density $n_{\rm e}$ ({\rm cm}$^{\rm-3}$), a
magnetic field $\textbf{B}$ ($\mu${\rm G}), and a path length $\textbf{l}$
({\rm kpc}), the Faraday RM is
\begin{equation}
\mathrm{RM}=812\int _{\rm 0}^{\rm l}n_{\rm e[{\rm cm}^{\rm -3}]}\textbf{B}_{\rm [\mu G]}\cdot \textbf{dl}_{\rm [kpc]}~~~\mathrm{rad}~\mathrm{m}^{\rm -2}.
\label{eq2}
\end{equation}

We produced the Faraday RM image by running the FARADAY code (Murgia
et al. 2004).  The software requires as input $Q$ and $U$ images for
each frequency and outputs the RM, the intrinsic
polarization angle $\Psi_{\rm int}$, their errors images, and
$\chi ^{\rm 2}$ maps. The RM image is created pixel by pixel by fitting the
observed polarization angle $\Psi_{\rm obs}$ versus the squared
wavelength $\lambda ^{\rm 2}$ for all the frequencies. To reduce
the problems associated with $n\pi$ ambiguities, the fitting algorithm
can perform a sequence of improvement iterations. In the first
iteration, only a subset of high signal-to-noise pixels is
considered. In the successive iterations, lower signal-to-noise pixels are
gradually included and the information from the previous iteration is
used to assist the fit of the $\lambda^{\rm 2}$ law.

We considered only those regions of the radio source where the total
intensity emission at 8415\,{\rm MHz} is above 5$\sigma_{\rm I}$.  Only
pixels with an uncertainty in polarization angle below 10$^{\rm \circ}$ at
each wavelength were considered. The RM fit is generated if this
condition is satisfied for at least five frequency maps. Almost half of
the total number of the pixels in the RM image are based on a
five-frequency fit. The remainder are based on a six-frequency fit.

\begin{figure*}[!]
   \centering
 \includegraphics[width=0.9\textwidth, angle=0]{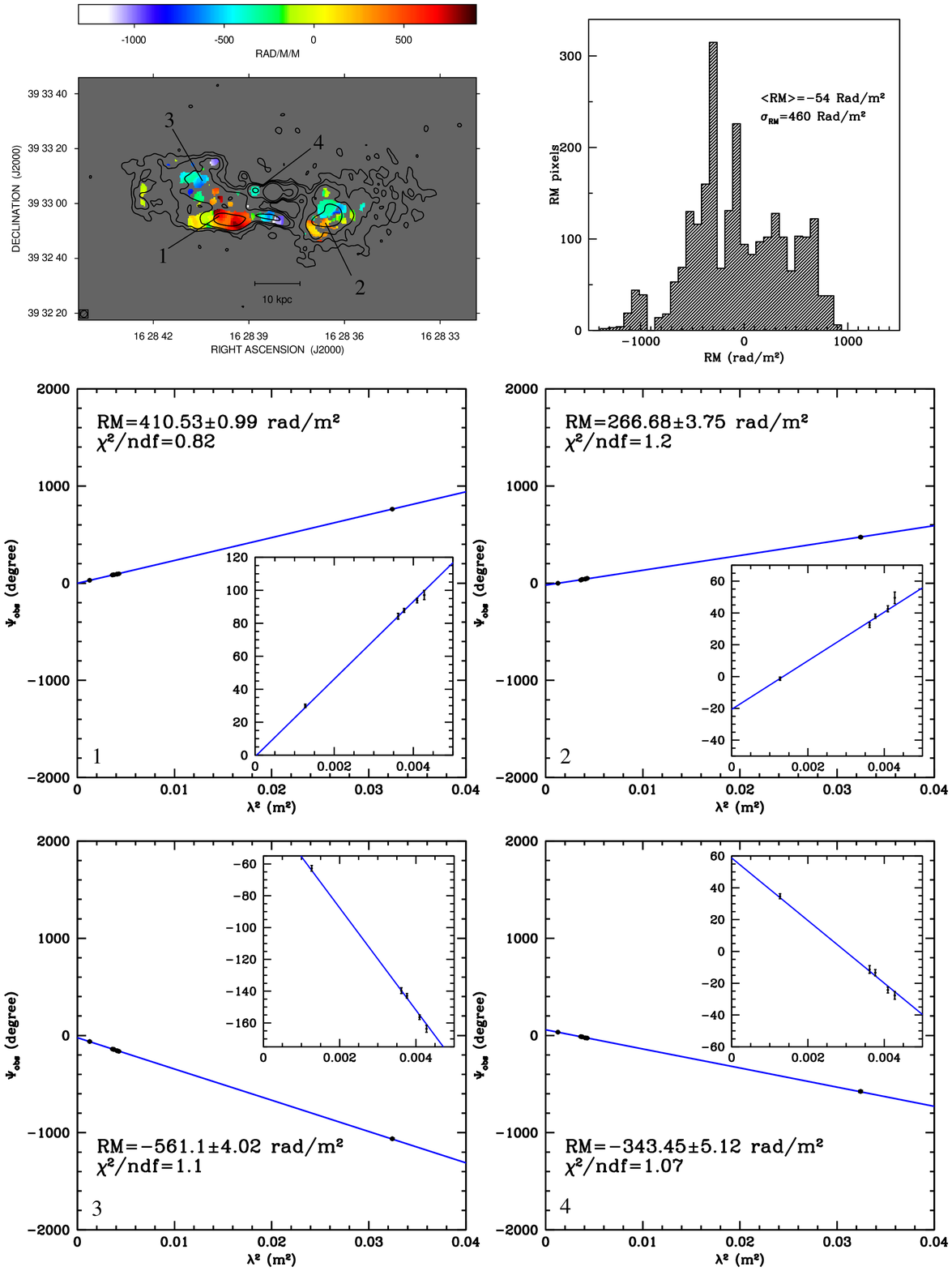}
      \caption{\emph{Top left}: total intensity radio contours of 3C\,338 at 8415\,{\rm MHz} 
        overlaid on the RM image. The angular resolution is $2.5\arcsec \times 2.5\arcsec$. 
        Contour levels are drawn at: 0.06, 0.12, 0.24, and 0.96\,mJy/beam. \emph{Top right}: 
        the histogram of the RM values. \emph{Bottom}: sample plots of the observed position angle 
        versus the squared wavelength at different source locations. The inset is a zoom of the 
        observed polarization angle corresponding to frequencies between 4585 and 8415\,{\rm MHz}.}
              \label{figure3}
    \end{figure*}

The final rotation measure image of 3C\,338 is shown in
Fig.\,\ref{figure3} (\emph{top left panel}) with total intensity contours at
8415\,{\rm MHz} overlaid. The image has a resolution of $2.5\arcsec$,
which corresponds to 1.5\,{\rm kpc} at the distance of 3C\,338. The RM has a
patchy structure with values ranging from $-$1300 to 900\,{\rm rad/m}$^{\rm 2}$. As
shown by the histogram in Fig.\,\ref{figure3} (\emph{top right panel}), the
RM distribution is characterized by a mean value of $\langle
$RM$\rangle=-54$\,rad/m$^{\rm 2}$ and a standard deviation $\sigma_{\rm
  RM}=460$\,rad/m$^{\rm 2}$. The RM fit is
characterized by a mean error $Err_{\rm fit}$=18\,rad/m$^{\rm 2}$. In Table\,\ref{table3} 
we report the statistics of the RM distribution of the entire source and, separately, of the 
east and west lobes. The mean and dispersion of the RM are quite different, but the ``total power'' 
$\langle RM^{\rm 2}\rangle^{\rm 1/2}=(\langle RM\rangle^{\rm 2}+\sigma_{\rm RM}^2)^{\rm 1/2}$ is very 
similar. Therefore, we cannot constrain the inclination to the line of sight of the source, and 
in the following analysis we assume that 3C\,338 is 
on the plane of the sky.  Examples of the
observed position angle $\Psi_{\rm obs}$ as a function of $\lambda^{\rm 2}$
for four high signal-to-noise pixels are shown in the \emph{bottom panels} of
Fig.\,\ref{figure3}. The observed fits are linear, as
expected in the case of a foreground Faraday screen.\\

\noindent
We have estimated the contribution of our own Galaxy to the Faraday
rotation in the direction of 3C\,338. In galactic coordinates, the
radio source is located at $l=62.9^{\rm \circ}$ and $b=43.7^{\rm \circ}$. We
computed the average of the RM values reported in Taylor et al. (2009)
for a region of about 3$^{\rm \circ}$ around this direction and from this 
estimated a Galactic contribution of $\sim$13 rad/m$^{\rm 2}$.  Since the
Galactic foreground is small compared to the RM intrinsic to 3C\,338,
it has a negligible impact on the RM.

\subsection{Depolarization}
\label{Burn law}
In Table\,\ref{fractional polarization} we report the average
fractional polarization levels of 3C\,338 
obtained by considering the same pixels as we used to calculate the RM
image. The radio source is less polarized at longer wavelengths. This
behavior can be interpreted in terms of variations is the RM on 
smaller scales than the beam of the radio images.

 \begin{table}
\caption{Fractional polarization between 1665 and 8415\,{\rm 
MHz}}             
\label{fractional polarization}      
\centering          
\begin{tabular}{c c}     
\hline\hline       
$\nu$&  FPOL \\
({\rm MHz})& (\%)              \\
\hline 
1665 & 1.1$\pm$0.3\\
4585 & 13.6$\pm$0.3\\
4685 & 15.8$\pm$0.2\\
4885 & 18.2$\pm$0.3\\
4985 & 23.0$\pm$0.4\\
8415 & 41.7$\pm$0.6\\
\hline    
 \multicolumn{2}{l}{\scriptsize Col. 1: Observation frequency;}\\ 
 \multicolumn{2}{l}{\scriptsize Col. 2: Fractional polarization.}\\ 
\end{tabular}
\end{table}
These unresolved RM structures in the foreground screen cause a depolarization of the signal 
which, to first order, can be approximated by  
\begin{equation}
FPOL=FPOL_{\rm 0} exp(-a\lambda^{\rm 4}),
\label{burn}
\end{equation}
where $a$ is a value related to the RM gradient within the observing
beam and $FPOL_{\rm 0}$ is the intrinsic fractional polarization (Burn 1966,
see also Laing et al. 2008 for a more recent derivation). We fitted
the Burn law only between 4585 and 8415\,{\rm MHz} (6 and 3.6\,{\rm cm}) observations, since the formula is not applicable in the
long-wavelength regime (20\,{\rm cm}/1665\,{\rm MHz}) (see
Tribble 1991 and \S \ref{2-dimensional simulations}). We find
$a$=(66$\pm$6)$\times 10^{\rm 3}$\,{\rm rad}$^2${\rm /m}$^{\rm 4}$.

\section{Faraday rotation modeling}
\label{Magnetic field modeling}

Our aim is to constrain the intracluster magnetic field
strength and structure in A2199 by using the information from the
radio galaxy RM and polarization images presented in the previous
section. Our modeling is based on the assumption that the Faraday rotation is
occurring entirely in the intracluster medium.  In particular, we 
suppose that there is no internal Faraday rotation inside the radio
lobes, as suggested by the X-ray cavities detected by Johnstone et
al. (2002) and by the observed linearity of the polarization angle
$\Psi_{\rm obs}$ versus $\lambda^{\rm 2}$ (see Fig.\,\ref{figure3}).  Moreover, we
suppose that any possible local RM enhancement occurring at the
interface between the radio lobes and the surrounding medium is
negligible compared to the total RM across the cluster (but see
Rudnick \& Blundell 2003 for a contrary viewpoint).

As shown in Eq.\,\ref{eq2}, the rotation measure is the integral along
the line-of-sight of the intracluster magnetic field and the thermal
gas density.  To derive useful information about the intracluster
magnetic field we first need a model for the spatial distribution of
the thermal electron density.

\subsection{Thermal gas modeling}
The distribution of the thermal electrons was modeled using the
\emph{Chandra} X-ray observation of A2199 by Johnstone et al. (2002).
\begin{figure*}[ht]
   \centering
 \includegraphics[width=1\textwidth]{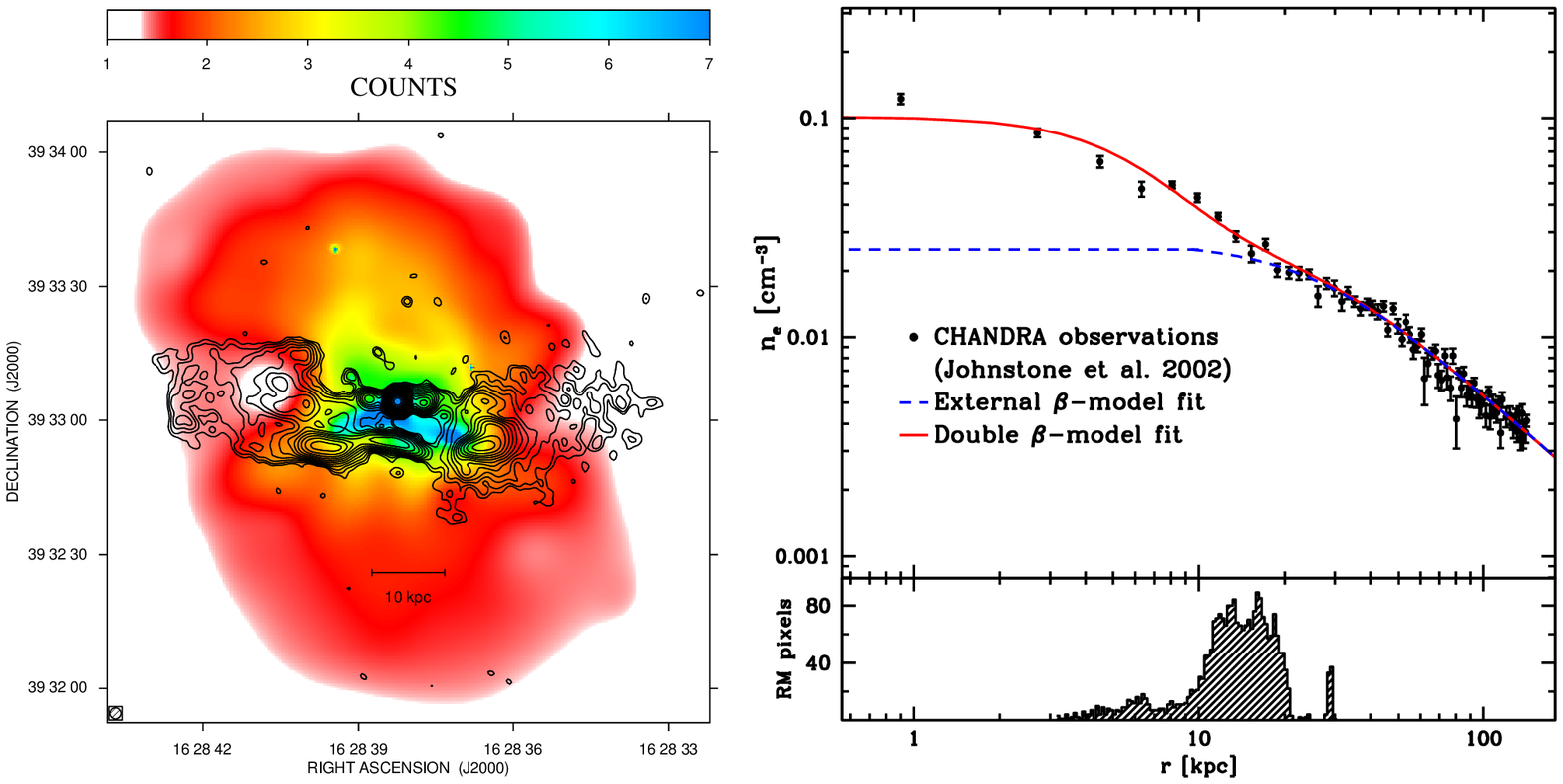}    
\caption{ \emph{Left panel}: Adaptively smoothed \emph{Chandra} X-ray image (Obs. ID  497) in the 0.2-10\,keV band of the galaxy cluster A2199 
       with the VLA contour levels at 8415\,{\rm MHz} overlaid. Pronounced X-ray cavities can be identified in correspondence 
       of the radio galaxy lobes. \emph{Top right panel}: deprojected thermal gas density profile 
       (Johnstone et al. 2002). \emph{Bottom right panel}: pixel distribution of the rotation 
         measure image as a function of the distance from the cluster center.}
              \label{figure8}
    \end{figure*}
In the \emph{left panel} of Fig.\,\ref{figure8} we show a superposition of
the VLA 8415\,{\rm MHz} contour levels on the \emph{Chandra} X-ray image in the 0.2-10\,{\rm keV} band of
the central region of A2199. In the \emph {top righthand panel} of Fig.\,\ref{figure8}
we show the deprojected thermal gas density $n_{\rm e}(r)$ profile derived by 
Johnstone et al. (2002). In the \emph{bottom righthand panel} we show the rotation measure pixel distribution 
as a function of the distance from the cluster center. The RM pixels are sampled 
at distances from the cluster core in the range 3 to 30\,{\rm kpc}, with a higher concentration between 10 and 20\,{\rm kpc}. 
 
We initially fit the deprojected thermal gas density profile with a
single $\beta$-model
\begin{equation}
n_{\rm e}(r)=n_{\rm 0}\left(1+\frac{r^{\rm 2}}{r_{\rm c}^{\rm 2}}\right)^{\rm -\frac{3}{2}\beta},
\label{eq5}
\end{equation}
where $r$ is the distance from the cluster X-ray centroid, $n_{\rm 0}$ the
central electron density, and $r_{\rm c}$ the cluster core radius
(Cavaliere \& Fusco-Femiano 1976). But this thermal gas model does not
allow a good description of the central jump owing to the high central
density, typical of cooling core galaxy clusters. Therefore, we tried
to improve the fit by considering a double $\beta$-model,

\begin{equation}
n_{\rm e}(r)=n_{\rm 0_{\rm int}}\left(1+\frac{r^{\rm 2}}{r_{\rm c_{\rm int}}^{\rm 2}}\right)^{\rm -\frac{3}{2}\beta_{\rm int}}+n_{\rm 0_{\rm  ext}}\left(1+\frac{r^{\rm 2}}{r_{\rm c_{\rm ext}}^{\rm 2}}\right)^{\rm -\frac{3}{2}\beta_{\rm ext}},
\label{doublebeta}
\end{equation}
where $n_{\rm 0_{\rm int}}$ and $n_{\rm 0_{\rm ext}}$ are the internal and
external central electron densities, while $r_{\rm c_{\rm int}}$ and
$r_{\rm c_{\rm ext}}$ are the internal and external cluster core
radii, respectively. We do not discuss the physical
validity of this model here. Our purpose is to empirically obtain an
analytic profile that provides a reasonable continuous description of
the deprojected gas density observed by Johnstone et al. (2002). This
analytic profile can be conveniently extrapolated to both the cluster
center and to large radii so that the integral in Eq.\,\ref{eq2} can
be performed.

The double $\beta$-model best-fit parameters are reported in
Table\,\ref{table4}, while the best-fit profile is shown as a continuous
line in the \emph{righthand panel} of Fig.\,\ref{figure8}. In the same plot we
also show, as a dotted line, the external part of the model that has
been constrained by a fit of the data points with $r>18$ {\rm kpc}. The
total central density we find is $n_{\rm 0}=n_{\rm 0_{\rm int}}+n_{\rm 0_{\rm
  ext}}$=0.101\,{\rm cm}$^{\rm -3}$.

\begin{table}
\caption{Double $\beta$- model parameters.}             
\label{table4}      
\centering          
\begin{tabular}{c c c}     
\hline\hline       
PARAMETER & VALUE & UNITS\\
\hline
$\beta_{\rm int}$ &1.5$^{\rm +0.2}_{\rm -0.5}$    &~\\
$r_{\rm c_{\rm int}}$   &9$^{\rm +2}_{\rm -3}$      &{\rm kpc}\\
$n_{\rm 0_{\rm int}}$   &0.074$^{\rm +0.004}_{\rm -0.01}$ &{\rm cm}$^{\rm -3}$\\
$\beta_{\rm ext}$ & 0.39$^{\rm +0.01}_{\rm -0.03}$ &~\\
$r_{\rm c_{\rm ext}}$   &26.0$^{\rm +0.8}_{\rm -6.0}$          &{\rm kpc}\\
$n_{\rm 0_{\rm ext}}$   &0.027$^{\rm +0.003}_{\rm -0.003}$  &{\rm cm}$^{\rm -3}$\\
\hline    
\end{tabular}
\end{table}

\subsection{Magnetic field modeling}
\label{3-dimensional simulations}

The patchy structure characterizing the RM image in Fig.\,\ref{figure3}
can be interpreted as a signature of the turbulent intracluster
magnetic field. In particular, the dispersion of the RM values can be
related to the strength and structure of the magnetic field. If the
magneto-ionic medium is approximated by uniform cells of size
$\Lambda_{\rm c}$ with random orientation in space, the Faraday rotation
from a physical depth $L\gg \Lambda_{\rm c}$ is expected to be a Gaussian
with zero mean and dispersion given by
\begin{equation}
\sigma_{\rm RM}^{\rm 2}=\langle RM^{\rm 2}\rangle = 812^{\rm 2}  \Lambda_{\rm c} \int_{\rm 0}^{\rm L} ( n_{\rm e} B_{\rm z} )^{\rm 2} dl,
\label{mono}
\end{equation}
where $B_{\rm z}$ is the magnetic field component along the line-of-sight (e.g., Felten 1996).

\begin{table*}
\caption{Magnetic field model parameters.}             
\label{table5}      
\centering          
\begin{tabular}{c c c} 
\hline\hline  
    Parameter         & Description    & Means of investigation\\ 
\hline                        
$\langle  B_{\rm 0}\rangle$      & strength at the cluster center & 3D simulations\\
  $\eta$                 & radial slope, $\langle  B(r)\rangle=\langle  B_{\rm 0}\rangle\left(\frac{n_{\rm e}(r)}{n_{\rm e}(0)}\right)^{\rm \eta}$& 3D simulations\\
  $n$                     & power spectrum index, $\vert B_{\rm k}\vert ^{\rm 2} \propto k^{\rm -n}$& 2D simulations\\
  $\Lambda_{\rm min}$         & minimum scale of fluctuation, $\Lambda_{\rm min}=2\pi/k_{\rm max}$& 2D simulations\\
  $\Lambda_{\rm max}$         & maximum scale of fluctuation, $\Lambda_{\rm max}=2\pi/k_{\rm min}$& 2D simulations\\
 \hline    
 \end{tabular}
\end{table*}
More generally, the magnetic field auto-correlation function $C_{\rm B_{\rm z}}(r)$ and the rotation 
measure auto-correlation function $C_{\rm RM}$ are related by the Abel transform (En{\ss}lin \& Vogt 2003),

\begin{equation}
C_{\rm RM}(r_{\rm \perp})=812^{\rm 2}n_{\rm e}^{\rm 2}L\times\left(2\int_{\rm r_{\perp}}^{\rm \sqrt{(L/2)^{\rm 2}+r_{\perp}^{\rm 2}}}\frac{C_{\rm B_{\rm z}}(r)r}{\sqrt{r^{\rm 2}-r_{\perp}^{\rm 2}}}dr\right),
\end{equation}
where, for simplicity, the thermal gas density $n_{\rm e}$ is considered constant here.

If the characteristic fluctuation scale of the magnetic field is much
smaller than the physical depth of the Faraday screen (such that $\langle
B \rangle=0$ and $\langle RM\rangle=0$), in the limit
$r_{\perp}\rightarrow 0$, we obtain

\begin{equation}
\sigma_{\rm RM}^{\rm 2}=812^{\rm 2}n_{\rm e}^{\rm 2}L\, \Lambda_{\rm B}\, \sigma_{\rm B_{\rm z}}^{\rm 2},
\end{equation}

\noindent
where

\begin{equation}
\Lambda_{\rm B}=\frac{2 \int_{\rm 0}^{\rm L/2} C_{\rm B_{\rm z}}(r) dr}{C_{\rm B_{\rm z}}(0)}
\end{equation}
defines the magnetic field auto-correlation length, which should be the appropriate 
scale to use for $\Lambda_{\rm c}$ in Eq.\,\ref{mono}.

For an isotropic divergence-free random field, it can be shown that $\Lambda_{\rm B}$ 
is related to the magnetic field power 
spectrum, $\vert B_{\rm k}\vert^{\rm 2}$, by 

\begin{equation}
\Lambda_{\rm B}=\frac{3\pi}{2}\frac{\int_{\rm 0}^{\rm \infty}\vert B_{\rm k}\vert^{\rm 2}kdk}{\int_{\rm 0}^{\rm \infty}\vert B_{\rm k}\vert^{\rm 2}k^{\rm 2}dk},
\end{equation}

\noindent
where $k=2\pi/\Lambda$ is the wave number (En{\ss}lin \& Vogt 2003).

This is a particularly important result, since it implies that we must know
the power spectrum of the magnetic field fluctuations to
determine the strength of the field from the RM image. We also note
that the RM power spectrum is proportional to the magnetic field power
spectrum,

\begin{equation}
\vert RM_{\rm k}\vert^{\rm 2} \propto n_{\rm e}^{\rm 2}L\, \vert B_{\rm k}\vert^{\rm 2},
\end{equation}
but the RM auto-correlation length,
\begin{equation}
\Lambda_{\rm RM}=2 \frac{\int_{\rm 0}^{\rm \infty}\vert RM_{\rm k}\vert^{\rm 2}dk}{\int_{\rm 0}^{\rm \infty}\vert RM_{\rm k}\vert^{\rm 2}kdk},
\end{equation}
is not the same as $\Lambda_{\rm B}$. 

To limit the number of free parameters, in this work we choose to
model a power-law power spectrum\footnote{Throughout this
  paper the power spectra are expressed as vectorial forms in
  $k$-space.  The one-dimensional forms can be obtained by multiplying
  by $4\pi k^{\rm 2}$ and $2\pi k$ the three and
  two-dimensional power spectra, respectively.}  with index $n$ of the form

\begin{equation}
\vert B_{\rm k}\vert ^{\rm 2}\propto k^{\rm -n}
\label{eq3}
\end{equation}

\noindent
in the wave number range from $k_{\rm min}$ to $k_{\rm max}$ and 0 outside. 
Moreover, we suppose that the power-spectrum normalization varies with the 
distance from the cluster center such that the average magnetic field strength 
scales as a function of the thermal gas density according to
\begin{equation}
\langle  B(r)\rangle=\langle  B_{\rm 0}\rangle\left[\frac{n_{\rm e}(r)}{n_{\rm 0}}\right]^{\rm \eta},
\label{eq4}
\end{equation}
where $\langle B_{\rm 0}\rangle$ is the average magnetic field strength at the center 
of the cluster, and $n_{\rm e}(r)$ is the thermal electron gas density. 
We have to note that the tapering of the magnetic field power spectrum due to the thermal gas density may not preserve exactly the magnetic-field power-spectrum power-law shape at the edges (see Fig.\,\ref{profile}).

Overall, our magnetic field model depends on the five parameters
listed in Table\,\ref{table5}: the strength at the cluster center
$\langle B_{\rm 0}\rangle$, the radial slope $\eta$, the power
spectrum index $n$, and finally the minimum and maximum scales of
fluctuation, $\Lambda_{\rm min}=2\pi/k_{\rm max}$ and $\Lambda_{\rm
  max}=2\pi/k_{\rm min}$, respectively.

\subsection{Bayesian inference}
\label{bayesian}
To constrain the magnetic field strength and structure, we proceeded
in two steps. First, we performed a two-dimensional analysis of the RM
fluctuations and of the source depolarization to constrain the slope
$n$ and the range of scales of the power spectrum. Second, we
performed three-dimensional numerical simulations to constrain the
strength of the field and its scaling with the gas density. In both
cases, we made use of the FARADAY code to produce synthetic
polarization images of 3C\,338 and to compare them to the observed
ones.  We compared model and data using the Bayesian inference, whose
use was first introduced in the RM analysis by En{\ss}lin \& Vogt
(2003). Because of the random nature of the intracluster magnetic
field, the RM image we observe is just one possible realization of the
data. Different realizations of magnetic field characterized by the
same power spectrum will generate different RM images. Thus, rather
than try to determine the particular set of power spectrum
parameters that best reproduces the given realization of the data, it
is perhaps more meaningful to search for that distribution of model
parameters that maximizes the probability of the model given the data. 
The Bayesian inference offers a natural
theoretical framework for this approach.

The Bayesian rule relates our prior information on the distribution
$P(\vec\theta)$ of model parameters $\vec\theta$ to their posterior
probability distribution $P(\vec\theta~\vert~D)$ after the data $D$
have been acquired

\begin{equation}
P(\vec\theta~\vert~D)=\frac{L(D~\vert~\vec\theta)P(\vec\theta)}{P(D)},
\label{bayes}
\end{equation}

\noindent
where $L(D~\vert~\vec\theta)$ is the likelihood function, while $P(D)$
is called the evidence. The evidence acts as a normalizing constant
and represents the integral of the likelihood function weighted by the
prior over all the parameters space
\begin{equation}
P(D)=\int L(D~\vert~\vec\theta)P(\vec\theta)d\vec\theta.
\end{equation} 

The most probable configuration for the model parameters is obtained
by maximizing the joint posterior given by the product of the
likelihood function with the prior probability.
We used a \emph{Markov Chain Monte Carlo} (MCMC) method to extract
samples from the posterior probability distribution. In particular, we
implemented the Metropolis$-$Hastings algorithm, which is capable of
generating a sample of the posterior distribution without the need to
calculate the evidence explicitly, which is often extremely difficult
to compute since it would require exploring the entire prior space.
The MCMC is started from a random initial value $\vec\theta_{\rm 0}$ and the
algorithm is run for many iterations by selecting new states according
to a transitional kernel, $Q(\vec\theta,\vec\theta^{\rm \prime})$, between
the actual, $\vec\theta$, and the proposed position,
$\vec\theta^{\rm \prime}$. The proposed position is accepted with
probability

\begin{equation}
h=\min \left[1, \frac{L(D~\vert~\vec\theta^{\rm \prime}) P(\vec\theta^{\rm \prime}) Q(\vec\theta^{\rm \prime},\vec\theta) } { L(D~\vert~\vec\theta) P(\vec\theta) Q(\vec\theta,\vec\theta^{\rm \prime}) } \right].
\end{equation} 

We chose for $Q$ a multivariate Gaussian kernel. The MCMC starts with
a number of ``burn-in'' steps during which, according to common
practice, the standard deviation of the transitional kernel is adjusted so
that the average acceptance rate stays in the range 25\%$-$50\%. After the
burn-in period, the random walk forgets its initial state and the chain
reaches an equilibrium distribution. The burning steps are discarded,
and the remaining set of accepted values of $\vec\theta$ is a
representative sample of the posterior distribution that can be used
to compute the final statistics on the model parameters.

\section{2D analysis}
\label{2-dimensional simulations}
We performed a preliminary two-dimensional analysis of the RM
fluctuations and of the source depolarization to constrain the slope
$n$ and the range of scales of the power spectrum.  The two-dimensional
analysis relies on the proportionality between the magnetic field and
the rotation measure power spectra. On the basis of this proportionality, 
the index $n$ of the two-dimensional rotation measure power spectrum is 
the same as the three-dimensional magnetic field power spectrum:
\begin{equation}
\vert RM_{\rm k}\vert ^{\rm 2}\propto k^{\rm -n}.
\label{rmpower}
\end{equation}

At this stage, we are only interested in the shape of the power
spectrum and not in the exact value of the normalization. Indeed,
as a first approximation, we do not consider the effect of the
spatially variable gas density; i.e., we suppose that the depth of
the Faraday screen is on average the same over all of the source.  We
simulated synthetic images with a given power spectrum in a
two-dimensional grid. The simulations start in the Fourier space, where
the amplitudes of the RM components are selected according to
Eq.\,\ref{rmpower}, while the phases are completely random. The RM
image in the real space is obtained by a fast Fourier transform (FFT).

As a compromise between computational speed and spatial dynamical
range, we adopted a computational grid of 2048$\times$2048\,pixels
with a pixel size of 0.05\,{\rm kpc}. This grid allowed us to explore RM
fluctuations on scales as small as $\Lambda_{\rm min}\simeq 0.1$\,{\rm kpc},
i.e. one order of magnitude smaller than the beam of the observations
($FWHM \simeq 1.5$\,{\rm kpc}). Simultaneously, we were able to investigate
fluctuations as large as $\Lambda_{\rm max}\simeq 100$\,{\rm kpc},
i.e., comparable to the linear size of 3C\,338.
 \begin{figure*}[!]
   \centering
   \includegraphics[width=1\textwidth]{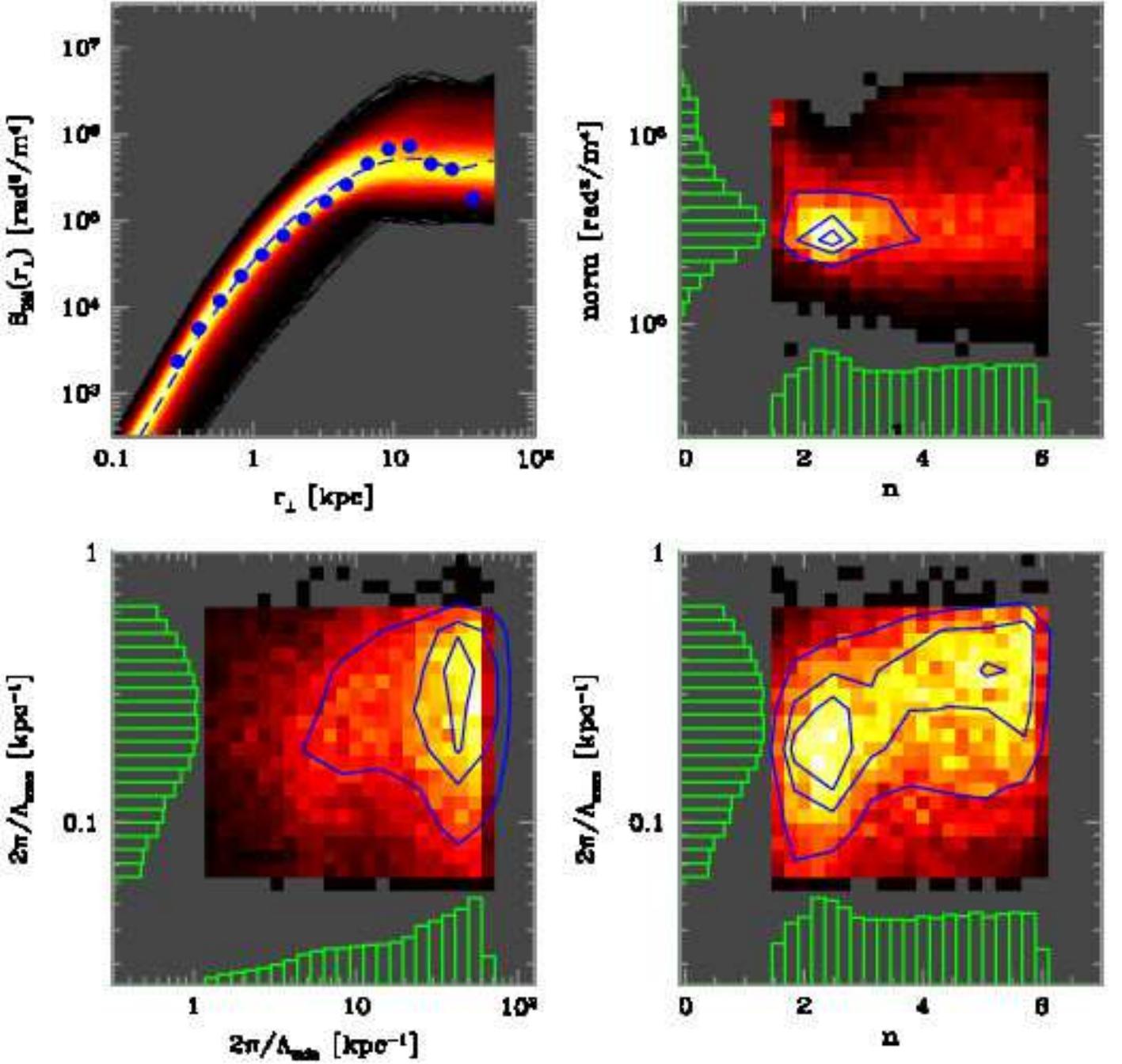}
\caption{\emph{Top left panel}: Bayesian analysis of the RM structure function. 
  The dots represent the data (error bars are comparable to the size of the symbols). 
  The shaded area represents the population of  synthetic RM structure functions from 
  the posterior distribution. The dashed line corresponds to the most probable value 
  for the model parameters (see text). \emph{Top right, bottom left, and bottom right panel}:  
  one-dimensional (histograms) and two-dimensional (colors and contours) marginalization of the 
  posterior for the model parameters. The contours are traced at 0.9, 0.75, and 0.5 of the
  peak value.}
             \label{structure_function_bayes}
    \end{figure*}
\begin{figure*}[!]
   \centering
   \includegraphics[width=1\textwidth]{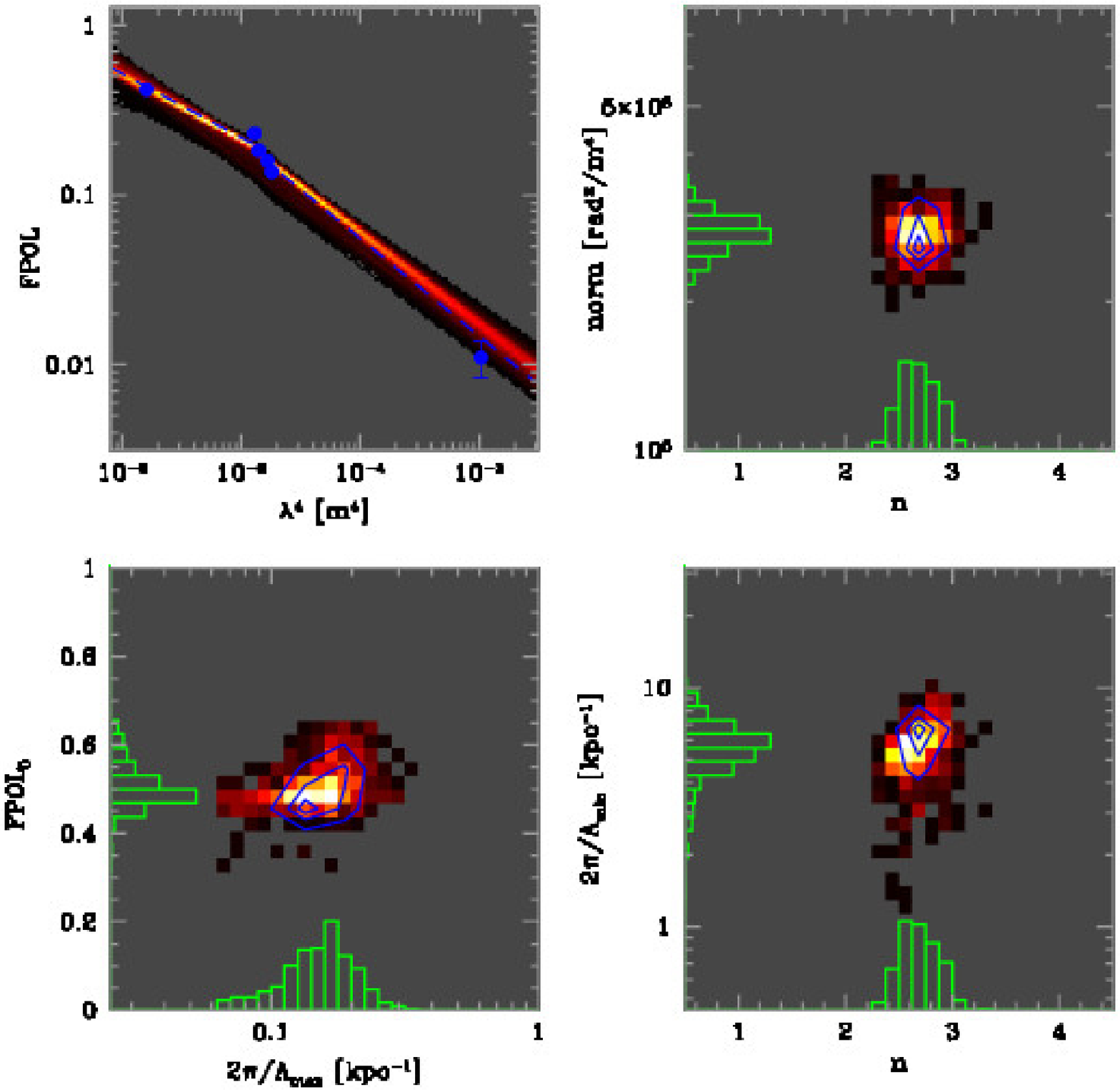}
\caption{\emph{Top left panel}: Bayesian analysis of the source depolarization. 
  The dots represent the data. The shaded area represents the population of synthetic 
  polarization from the posterior distribution. The dashed line corresponds to the most 
  probable value for the model parameters (see text). 
  \emph{Top right, bottom left, and bottom right panel}:  one-dimensional (histograms) 
  and two-dimensional (colors and contours) marginalization of the posterior for the 
  model parameters. The contours are traced at 0.9, 0.75, and 0.5 of the
  peak value.}
              \label{depolarization_lmin}
    \end{figure*}
We determined the distribution of the best-fit parameters using the
Bayesian approach outlined in \S\ref{bayesian}.  In particular, we
characterize the RM image by its structure function
\begin{equation}
S_{\rm RM}(r_{\rm \perp})=\langle [RM(x,y)-RM(x+\Delta x, y+\Delta y)] ^{\rm 2}\rangle_{\rm (x,y)},
\end{equation}
which is obtained by averaging the difference in RM values corresponding to pixels
located at the scale distance $r_{\rm \perp}=\sqrt{\Delta x^{\rm 2}+\Delta
  y^{\rm 2}}$, and is related to the RM auto-correlation function
\begin{equation}
C_{\rm RM}(r_{\rm \perp})=\langle RM(x,y)\cdot RM(x+\Delta x, y+\Delta y)\rangle_{\rm (x,y)}
\end{equation}
by the simple relation
\begin{equation}
S_{\rm RM}(r_{\rm \perp})=2(\sigma_{\rm RM}^{\rm 2}+\langle RM\rangle^{\rm 2})-2C_{\rm RM}(r_{\rm \perp}).
\end{equation}
The observed structure function is shown in the \emph{top lefthand panel} of
Fig.\,\ref{structure_function_bayes}, the formal error bars are
comparable to the size of the dots. For an isotropic field, the RM
auto-correlation function and power spectrum are related by the Hankel
transform
\begin{equation}
C_{\rm RM}(r_{\rm \perp})= 2\pi \int_{\rm 0}^{\rm \infty} J_{\rm 0}(kr_{\rm \perp}) \arrowvert RM_{\rm k}\arrowvert ^{\rm 2} k dk,
\label{hankel}
\end{equation}
where
\begin{equation}
J_{\rm 0}(kr_{\rm \perp})=\frac{1}{2\pi}\int_{\rm 0}^{\rm 2\pi}e^{\rm -ikr_{\rm \perp}cos\theta}d\theta
\end{equation}
is the zero-order Bessel function. For the power-law power spectrum in
Eq.\,\ref{rmpower}, we can identify three regimes in the RM structure
function. In the asymptotic small-separation regime, $r_{\rm \perp}\ll
2\pi/k_{\rm max}$, the structure function increases as
$S_{\rm RM}(r_{\rm \perp})\propto r_{\rm \perp}^{\rm 2}$. In the intermediate regime,
where $2\pi/k_{\rm max} \ll r_{\rm \perp}\ll 2\pi/k_{\rm min}$,
$S_{\rm RM}(r_{\rm \perp})\propto r_{\rm \perp}^{\rm n-2}$. Finally for $r_{\rm \perp}\gg
2\pi/k_{\rm min}$, the structure function saturates to the constant value
of $S_{\rm RM}\simeq 2\sigma_{\rm RM}^{\rm 2}$. However, it can be hard to
discern these three regimes directly from the observed structure
function because of the coarse resolution of the radio images and the
undersampling of the large separations because of the finite size of the
RM image. Indeed, we need to resort to numerical simulations in order
to account for the effects of these window functions. The
synthetic images are thus gridded to the same geometry as the data and are
convolved to the same angular resolution. Moreover, we masked the
synthetic images using the observations to reproduce the window function
imposed by the shape of 3C\,338, and we added Gaussian noise with an
rms value of $Err_{\rm fit}$, in order to mimic the noise of the
observed RM image.
\begin{figure*}[ht]
   \centering
  \includegraphics[width=15cm]{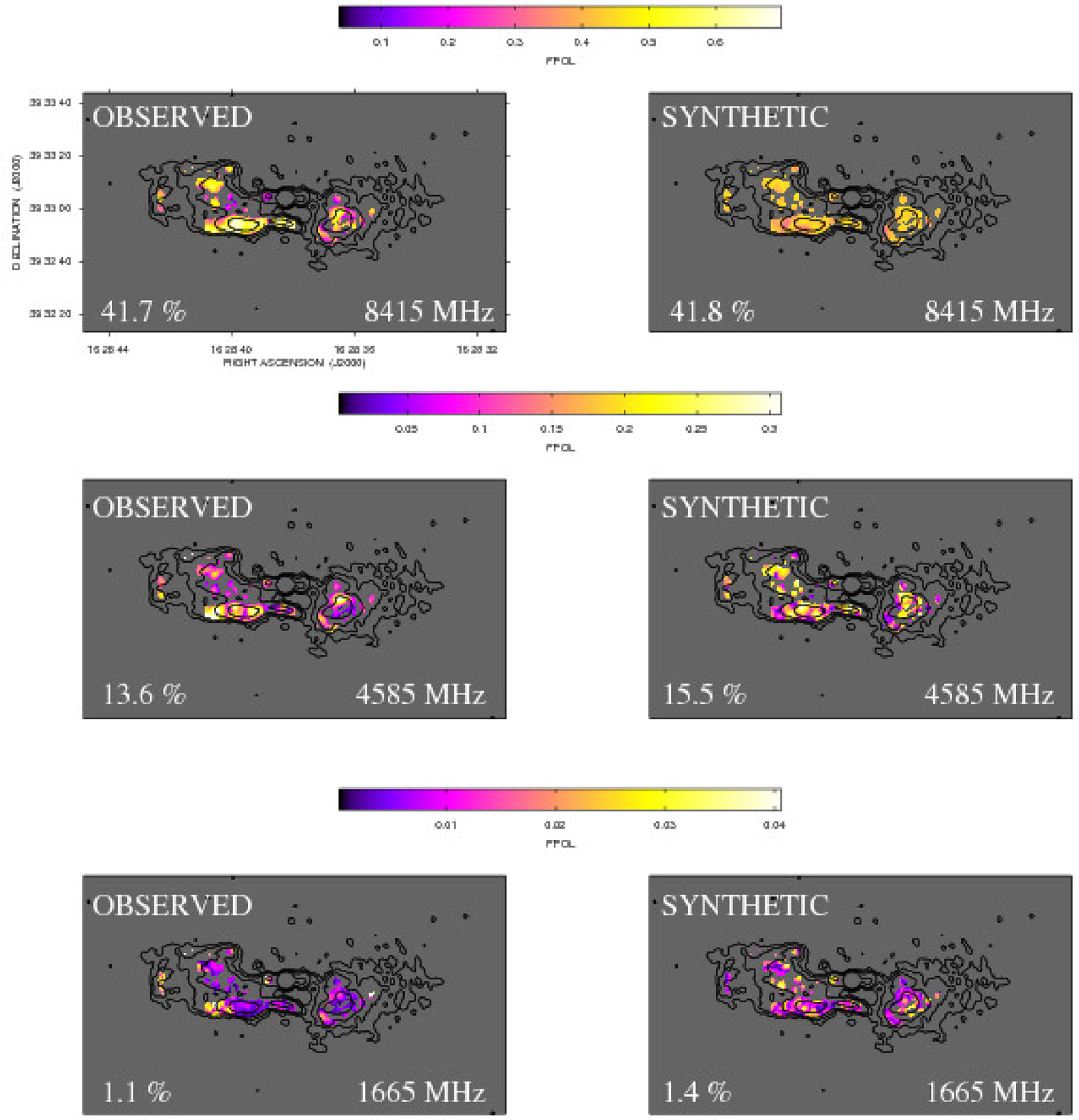}
      \caption{Examples of observed (\emph{left}) fractional polarizations images 
        at 8415, 4585, 1665\,{\rm MHz} compared with synthetic realizations (\emph{right}) 
        corresponding to the best fit parameters from the depolarization analysis.}
              \label{Polarization comparison}
    \end{figure*}

We then applied the Bayesian method described in \S\,\ref{bayesian} by
choosing uniform priors for the four power spectrum parameters: the normalization
$norm$, the slope $n$, the minimum, and the maximum wave number 
$k_{\rm min}$ and $k_{\rm max}$.  The statistics of the
simulated RM fluctuations are Gaussian, so their structure
function at a given separation has a log-normal distribution. At each
step of the MCMC, we performed 30 realizations of the same power
spectrum with different random seeds and evaluated the average and
the dispersion of the logarithm of the model structure function. 
We averaged the offsets and the dispersions at all the individual separations. 
From these values we computed the likelihood of the observed structure function.

In Fig.\,\ref{structure_function_bayes} we show the results of the
Bayesian analysis of the rotation measure structure function. The
output from the MCMC is a 4-dimensional hypercube containing a sample
of the posterior distribution of the power spectrum
parameters. The \emph{top-right, bottom-left, and bottom righthand panels} show the
two-dimensional marginalization of this hypercube as color images. In
addition, one-dimensional marginalizations are shown as histograms along
each axis of these images. The two-dimensional marginalizations
represent the projected density of samples in the MCMC, which is
proportional to the posterior probability of that particular couple of
model parameters.  The maximum scale of the fluctuation, the
normalization, and the slope of the power spectrum appear to be
characterized by a peak that corresponds to the maximum posterior
probability for that configuration of parameters, while for the
minimum scale of fluctuation we only have an upper limit.

To provide a visual comparison between model and data, in the \emph{top lefthand
panel} of Fig.\,\ref{structure_function_bayes} we show the observed
structure function along with the population of synthetic
structure functions contained in the posterior sample and 
the structure function for which posterior is
maximum (best fit). The brighter
pixels in the shaded image occur where many different synthetic
structure functions overlap each other. The probability of the model 
given the data is higher in the
brighter regions and is lower in the darker regions. The shaded region
is relatively narrow for small separations but widens significantly
for $r_{\rm \perp} > 10$\,{\rm kpc}, indicating that our sensitivity to the
large-scale separations is lower.  Overall, the data stay close to the
high-probability region, indicating that the model is doing a good job
reproducing the observed RM structure. However, for $r_{\rm \perp} > 20$\,{\rm 
kpc} the observed structure function decreases while the model stays
relatively constant. The turnover of the observed structure function
is very likely due to the lower gas density at large distances from the
cluster center, which results in a systematic decrease in the power
spectrum normalization, hence of the Faraday rotation on large
scales. This effect is not included in the current modeling, but
it will be investigated with the aid of three-dimensional simulations in
\S\ref{The magnetic field model}.
 
The power spectrum used to model the RM image should also be
consistent with the observed depolarization of the radio source at
increasing wavelengths (see \S\ref{Burn law}). The depolarization is
caused by RM variations on smaller scales than the beam that results
in an incoherent sum of the radio signal. Indeed, modeling of the
polarization amplitude can be used to place more stringent constraints
on the minimum scale of fluctuation of the magnetic field and on the
slope of the power spectrum. Using the same simulation set-up
presented above, we reproduced the expected polarized signal from
3C\,338 as a function of the wavelength. We first constructed an image
of the source polarization at $\lambda=0$, which is characterized by
an intrinsic degree of polarization $FPOL_{\rm 0}$. We then simulated
different RM images at full resolution and used them to rotate the
intrinsic polarization vectors according to Eq.\,\ref{eq1}. These
full-resolution images are finally convolved at the same resolution as
the observations, resulting in beam depolarization of the signal at
longer wavelengths.  

We used the Bayesian inference and a Gaussian
likelihood to estimate the distribution of the power spectrum
parameters, which maximizes the probability that the observed
depolarization is a realization of the model.  In
Fig.\,\ref{depolarization_lmin} the posterior from the depolarization
analysis is shown for the five free parameters: the intrinsic degree
of polarization $FPOL_{\rm 0}$, the normalization $norm$, the slope $n$, the
minimum, and the maximum scale of fluctuation $\Lambda_{\rm min}$, and
$\Lambda_{\rm max}$ of the magnetic field power spectrum.  In the \emph{top right and bottom panels} the two-dimensional (colors and contours)
and one-dimensional (histograms) marginalization of the posterior are
shown for three different combinations of the model parameters. All
the model parameters appear to be well constrained, and their values
are consistent with the structure function analysis.  In the \emph{top lefthand
panel}, the observed fractional polarization as a function of the
fourth power of the wavelength is shown, together with the sample of
realizations from the posterior and the best fit.
In Fig.\,\ref{Polarization comparison} the observed fractional
polarization images are compared with synthetic realizations
corresponding to the best fit parameters.  The synthetic fractional
polarization trend at higher frequencies (4585 to 8415\,{\rm MHz}) is
consistent with a Burn law, $FPOL=FPOL_{\rm 0}\exp(-a\lambda^{\rm 4})$, with
$a_{\rm synth}$=(61$\pm$3)$\times$10$^{\rm 3}$rad$^{\rm 2}$/m$^{\rm 4}$, in very good
agreement with the observed value $a_{\rm
  obs}$=(66$\pm$6)$\times$10$^{\rm 3}${\rm rad}$^{\rm 2}${\rm /m}$^{\rm 4}$. However, it is clear from 
the \emph{top lefthand panel} of Fig.\,\ref{depolarization_lmin} that
our simulations are also able to explain the observed
polarization levels reasonably well at low frequencies (1400\,{\rm MHz}) where the Burn
law breaks down.

Overall, the combined two-dimensional analysis, RM image, and
depolarization allowed us to obtain a first insight into the shape of
the magnetic field power spectrum. We constrained the power spectrum
index to $n$=(2.8$\pm$1.3), and the minimum and maximum scales in the
range from $\Lambda_{\rm min}$=(0.7$\pm$0.1)\,{\rm kpc} to
$\Lambda_{\rm max}$=(35$\pm$28)\,{\rm kpc}, where the given errors represent 
the dispersion of the one-dimensional 
marginalizations. In the next step we fix these values
and constrain the strength of the magnetic field and its scaling
with the gas density with the aid of three-dimensional simulations.

\section{3D simulations}
\label{The magnetic field model}

\begin{figure}[h]
  \centering
   \includegraphics[width=9cm]{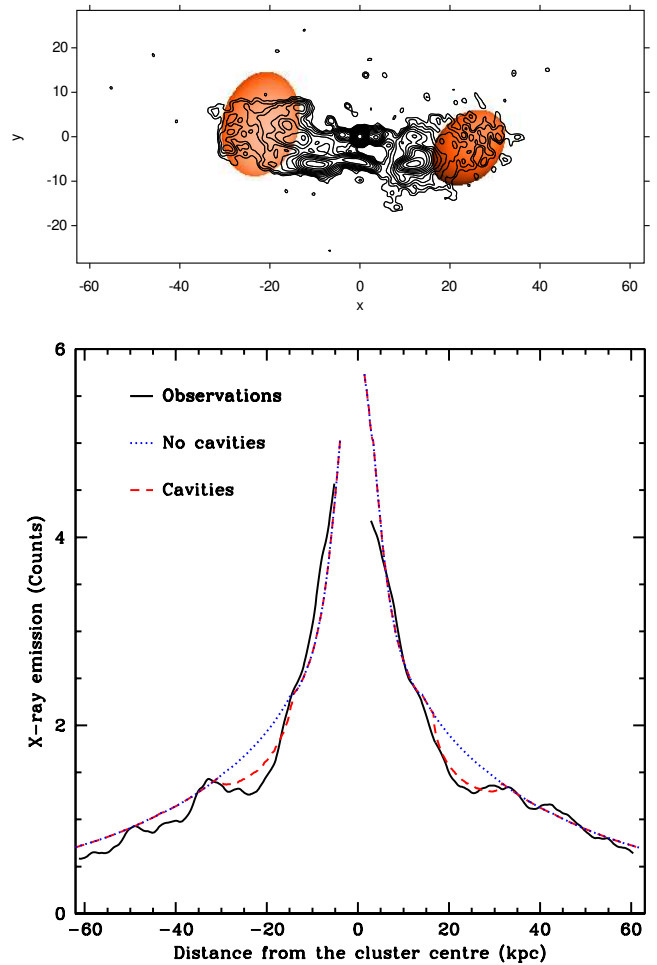}
  \caption{\emph{Top panel}: X-ray cavity model. The darker region indicates the location 
    where the thermal gas has been subtracted from the double $\beta$-model. \emph{Bottom panel}: 
    horizontal slice passing through the X-ray cavities at the cluster center. 
    Continuous line represents the observations from the \emph{Chandra} image, dashed line simulations 
    with cavities, dotted line simulations without cavities. }
  \label{cavities}
\end{figure}

We can construct a three-dimensional model of the intracluster magnetic
field by following the numerical approach described in Murgia et
al. (2004).  The simulations begin in Fourier space by extracting the
amplitude of the magnetic field potential vector, $\tilde A(k)$, from
a Rayleigh distribution whose standard deviation varies with the wave
number according to $|A_{\rm k}|^{\rm 2}\propto k^{\rm -n-2}$.  The phase of the
potential vector fluctuations is taken to be completely random. The
magnetic field is formed in Fourier space via the cross product
$\tilde B(k)=i k \times \tilde A(k)$. This ensures that the magnetic
field is effectively divergence free.  We then perform a three-dimensional 
FFT inversion to produce the magnetic field in the
real space domain.  The field is then Gaussian and isotropic, in the
sense that there is no privileged direction in space for the magnetic
field fluctuations. The power-spectrum normalization is set such that
the average magnetic field strength scales as a function of the
thermal gas density according to Eq.\,\ref{eq4}. The gas density
profile is described by the double-beta model in Eq.\,\ref{doublebeta}.
Actually, we know that the gas density distribution in A2199 deviates from 
spherical symmetry because of the X-ray cavities corresponding to
the radio lobes of 3C\,338. Hence, we 
include the X-ray cavities in the three-dimensional simulations by removing 
from the gas density modeling all the magneto-ionic material inside two 
ellipsoidal regions centered on the radio lobes, as shown in the 
\emph{top panel} of Fig.\,\ref{cavities}. In the \emph{bottom panel} we trace
the profile of the X-ray brightness along an horizontal slice passing
through the radio lobes for the double beta-model with and without
cavities. The model with cavities provides a better description of the
observed X-ray brightness along the slice.

We simulated the random magnetic field by using a cubical grid of
1024$^{\rm 3}$\,pixels with a cell size of 0.16\,{\rm kpc}/pixel. The synthetic 
RM images were obtained by integrating numerically the gas density and the
magnetic field product along the line-of-sight accordingly to
Eq.\,\ref{eq2}. In a similar way to the two-dimensional simulations,
the synthetic RM images were gridded, convolved, blanked, and noised as the 
observed RM image before the comparison with the data. We
used the Bayesian approach to find the posterior distribution for
$\langle B_{\rm 0}\rangle$ and $\eta$, which maximizes the probability that the
observed structure function is a realization of the model. The slope
$n$ and the scales $k_{\rm min}$ and $k_{\rm max}$ were kept fixed at
the values found with the two-dimensional analysis.

\label{X-ray cavities modeling}

\begin{figure}[hb]
  \centering
  \includegraphics[width=9cm]{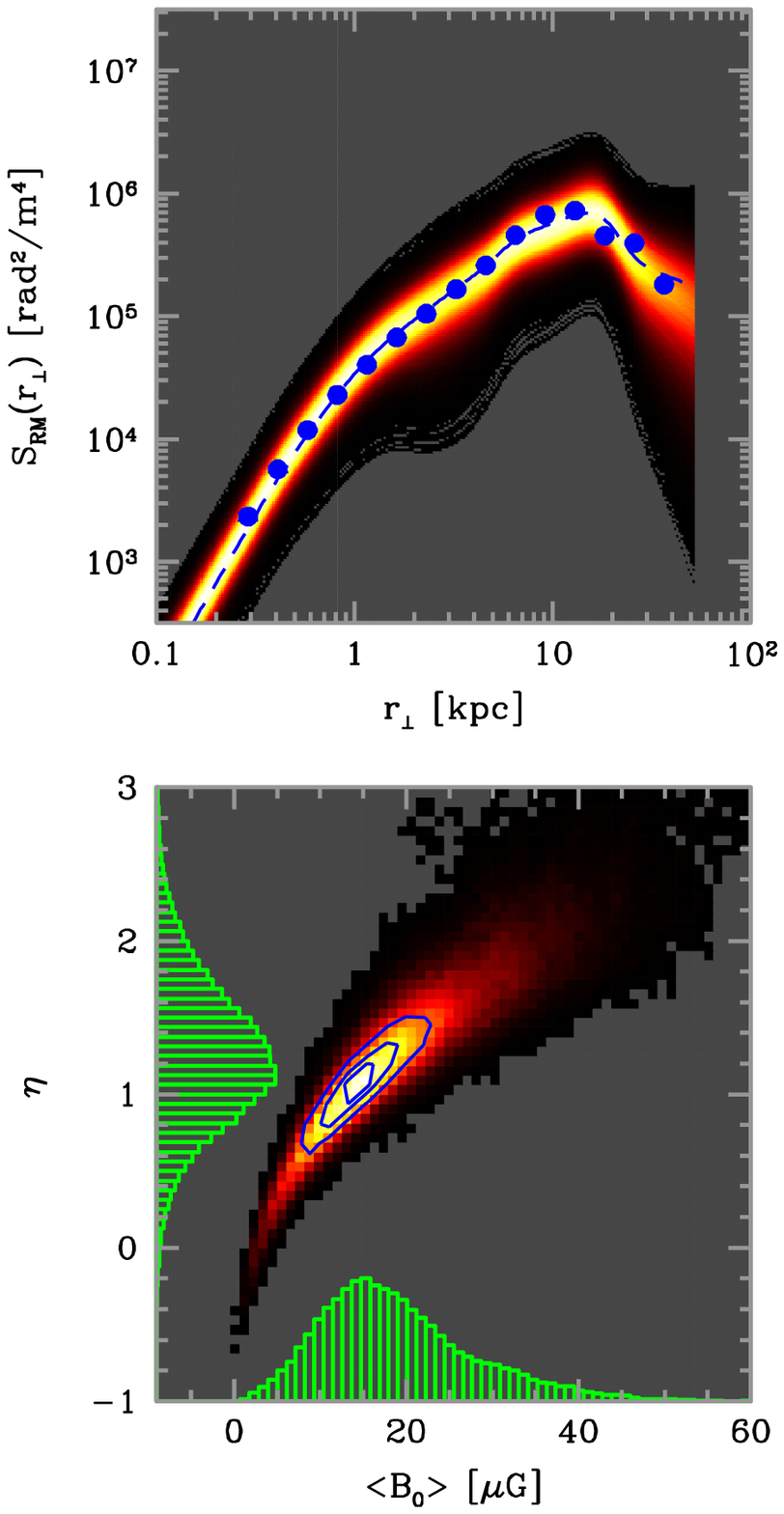}
  \caption{
    Bayesian 3-dimensional analysis of the RM structure function for the model with $\Lambda_{\rm max}$=35 {\rm kpc}.  \emph{Top panel}: The dots represent the data (error bars are comparable to the 
    size of the symbols). The shaded area represents the population of  synthetic RM structure functions 
    from the posterior distribution. The dashed line corresponds to the most probable value for the 
    model parameters (see text). \emph{Bottom panel}:  one-dimensional (histograms) and two-dimensional 
    (colors and contours) marginalizations of the posterior for the model parameters. The contours are 
    traced at 0.9, 0.75, and 0.5 of the
    peak value.}
  \label{B cavities}
\end{figure}

\begin{figure*}[h!]
  \centering
  \includegraphics[width=16cm]{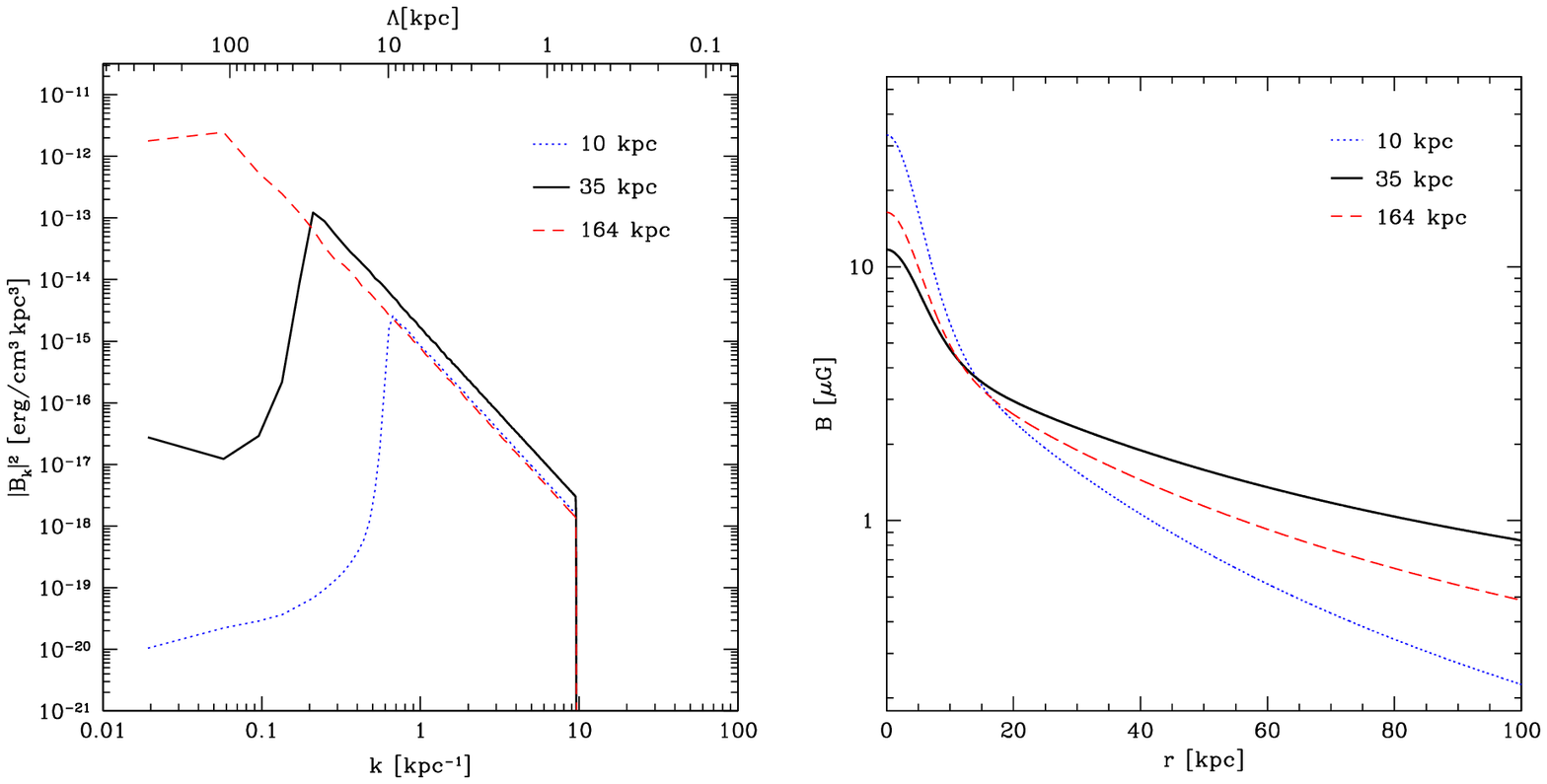}
  \caption{
 The magnetic field power spectra (\emph{left panel}) and radial profiles (\emph{right panel}) for the models corresponding 
to the three values of $\Lambda_{\rm max}$=10, 35, and 164 kpc are shown as dotted, continuous, and dashed lines, respectively.
The model with $\Lambda_{\rm max}$=35 kpc provides the best description of the observed RM structure function, see text. }
  \label{profile}
\end{figure*}

We started with uniform priors for $\langle B_{\rm 0}\rangle$ and $\eta$ and we
evaluated the likelihood of the structure function by mean of 30
different configurations for the magnetic field phases at each step of
the MCMC. The result of the Bayesian analysis is shown in Fig.\,\ref{B
  cavities}.  In the \emph{bottom panel} we present the two-dimensional
(colors and contours) and one-dimensional marginalizations
of the posterior. The two parameters appear well constrained.  We
found a magnetic field with a central strength $\langle B_{\rm
  0}\rangle$=(11.7$\pm$9.0)\,$\mu${\rm G}, and a radial slope
$\eta$=(0.9$\pm$0.5).  In the \emph{top panel} we show the observed
structure function  along with the synthetic structure functions
from the posterior and the best fit. The three-dimensional
modeling represents a significant improvement over the
two-dimensional analysis. In fact, we are now able to describe the overall shape of the observed structure function with good accuracy,
including the turnover at large separations that is most likely due
to the decrease in the Faraday rotation with radius. In the three-dimensional simulations we kept
the maximum scale of the magnetic field fluctuation fixed to
 $\Lambda_{\rm max}$=35\,{\rm kpc}, which is the value found from the two-dimensional
 analysis in \S\,\ref{2-dimensional simulations}. The consequences of a different
 choice of this parameter are discussed in further detail in the Appendix where we also tested the cases $\Lambda_{\rm max}$=10 and 164 {\rm kpc}. The value
  $\Lambda_{\rm max}$=35\,{\rm kpc} still provides the best description of the 
observed RM structure function. In Fig.\,\ref{profile} we plot the magnetic field power spectra and radial profiles
 corresponding to the maximum posterior for the three values of $\Lambda_{\rm max}$.
 
We performed the same Bayesian analysis also not taking 
of the X-ray cavities into account (not shown). We found very similar results to those 
 including the cavities 
and we conclude that the presence of these voids
in the gas density distribution has a second-order impact on the Faraday rotation
measures we are analyzing. This could be because  
the most of the lines of sight sampled from our observed RM image do not 
intercept the regions of the cluster affected by the cavities.

\begin{figure*}[ht!]
  \centering
  \includegraphics[width=15cm]{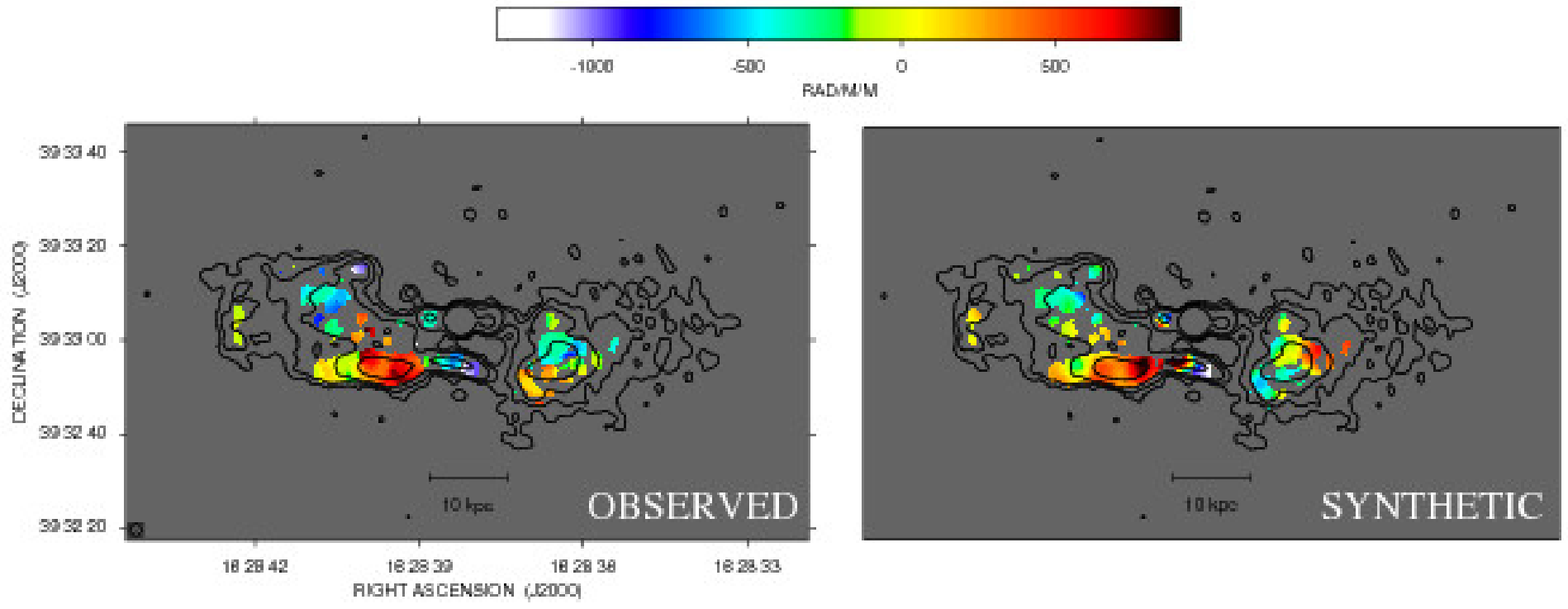}
  \caption{Qualitative comparison between the observed rotation measure image (\emph{left}) 
    and a synthetic 3-dimensional realization (\emph{right}).}
  \label{rm_comparison}
\end{figure*}

Finally, in Fig.\,\ref{rm_comparison} we present a qualitative
comparison between the observed and {\rm a} synthetic RM images taken from
the simulations with cavities. Although very simple, the power-spectrum model adopted here provides a reasonable description of the
patchy RM structure seen in the data.

It is interesting to compare our result with the independent estimate
based on the theoretical work by Kunz et al. (2011).  By assuming that
turbulent dissipation balances radiative cooling at all radii inside
the cluster core, Kunz et al. (2011) find in the bremsstrahlung
regime (that is $T\gtrsim$ 1\,keV)
\begin{equation}
B(r)\simeq 11\xi^{\rm -1/2}\left(\frac{n_{\rm e}(r)}{0.1{\rm cm}^{\rm -3}}\right)^{\rm 1/2}\left(\frac{T(r)}{2{\rm keV}}\right)^{\rm 3/4} \mu {\rm G},
\label{kunz}
\end{equation}
where $T$ is the temperature, while $\xi$ is expected to range between
0.5 and 1 in a turbulent plasma.  By considering a central density $n_{\rm 0}$$\simeq$0.1\,{\rm cm}$^{\rm -3}$ and a
central temperature $T$$\simeq$2\,keV, according Johnstone et al. (2002), they found $B_{\rm
  0}$$\simeq$11$\xi^{\rm -1/2}$\,$\mu${\rm G}, very close to our estimate based
on the Faraday rotation.

\section{Conclusions}
\label{Conclusions}

We investigated the magnetic field power spectrum in the cool core
galaxy cluster A2199 by analyzing the polarized emission of the
central radio source 3C\,338. We used archival VLA observations 
between 1665 and 8415\,{\rm MHz} to produce detailed Faraday rotation measure and
fractional polarization images of the radio galaxy. 
We observed a significant depolarization of the radio emission of the radio galaxy 
and high RM values. This agrees with the analysis performed by 
Ge\& Owen (1994) at 5000\,{\rm MHz}, although our result takes the 
additional information at 1665 and 8415\,{\rm MHz} into account.

 We simulated
Gaussian random three-dimensional magnetic field models with different
power-law power spectra, and we assumed that the field strength
decreases radially as a power of the thermal gas density as
$n_{\rm e}^{\rm \eta}$. By comparing the synthetic and the observed images with
a Bayesian approach, we constrained the strength and structure of the
magnetic field associated with the intracluster medium.
We found that the Faraday rotation toward 3C\,338 in A2199 is
consistent with a magnetic field power-law power spectrum
characterized by an index $n$=(2.8$\pm$1.3) between a maximum and a
minimum scale of fluctuation of $\Lambda_{\rm max}$=(35$\pm$28)\,{\rm kpc} and $\Lambda_{\rm min}$=(0.7$\pm$0.1)\,{\rm kpc}, respectively. 
The corresponding magnetic field auto-correlation length is $\Lambda_{\rm B}$=5.2\,{\rm kpc}.
In addition, by including in the modeling the presence of X-ray cavities in
coincidence with the radio galaxy lobes, we found a magnetic field strength of $\langle B_{\rm
  0}\rangle$=(11.7$\pm$9.0)\,$\mu${\rm G} at the cluster center. Farther out, the field decreases with radius following the gas density to the power of $\eta$=(0.9$\pm$0.5).
To a comparison with other values reported in the literature, 
the radially averaged magnetic field
strength calculated over the central 1\,{\rm Mpc}$^{\rm 3}$ is $\sim$0.19\,$\mu${\rm G}. 
The additional data and the numerical modeling of the intracluster magnetic field 
fluctuations allowed us to improve upon the previous estimate of 15\,$\mu${\rm G} 
(Eilek \& Owen 2002), with more stringent constraints not only on the magnetic field 
strength but also on its structure.

From the literature the central magnetic field strength found 
 in merger galaxy clusters is a few $\mu${\rm G} 
(e.g., 2.5\,$\mu${\rm G} in A2255 by Govoni et al. 2006 and 
4.7\,$\mu${\rm G} in Coma by Bonafede et al. 2010). In cooling-core galaxy clusters 
the magnetic field central strength is still uncertain. 
Typically it is a few tens $\mu${\rm G} (as in Hydra\,A, where values between 19 and 
80\,$\mu${\rm G} have been indicated, e.g. 
Laing et al. 2008 and Kuchar \& Ensslin 2011), even if lower values have also been found, as in the galaxy 
cluster  A2634 (3\,$\mu${\rm G}, Vogt \& Ensslin 2003). In A2199 we found a magnetic field strength 
of about 10\,$\mu${\rm G} that seems to confirm values of the magnetic field 
central 
strength in cooling-core galaxy cluster higher than in the merging cluster.

\begin{acknowledgements}
We thank the anonymous referee 
for the useful comments fundamental for improving this work. We are grateful to A. Bonafede for useful discussions.
This work made use of results produced by the Cybersar Project
managed by the Consorzio COSMOLAB, a project cofunded by the Italian
Ministry of University and Research (MIUR) within the Programma
Operativo Nazionale 2000-2006 ``Ricerca Scientifica, Sviluppo
Tecnologico, Alta Formazione'' per le Regioni Italiane dell' Obiettivo
1 (Campania, Calabria, Puglia, Basilicata, Sicilia, Sardegna) -  Asse
II, Misura II.2 ``Societ\`a dell$'$Informazione'', Azione a ``Sistemi di
calcolo e simulazione ad alte prestazioni''. More information is
available at http://www.cybersar.it.  This research was partially
supported by PRIN-INAF2009. The National Radio Astronomy Observatory (NRAO) is a
facility of the National Science Foundation, operated under
cooperative agreement by Associated Universities, Inc. We are grateful
to Antonella Fara for the assistance with the
Cybersar-OAC computer cluster.
\end{acknowledgements}

\appendix
\begin{onecolumn}
\section{Maximum scale of fluctuation}
\label{sec:app}

In this appendix we briefly discuss the determination of the outer scale of
the magnetic field fluctuations. On the basis of the two-dimensional and 
 three-dimensional simulations, we find that a maximum fluctuation scale of $\Lambda_{\rm max}$=35\,kpc 
provides a very good description of the observed structure function.

The observed rotation measure structure function is affected by a turnover in 
the large-separation regime ($r_{\rm \perp}\gg2\pi/k_{\rm min}$). The drop
comes from a lowering of the gas density at large distances from the
cluster center, resulting in a systematic decrease of the power of the Faraday rotation on large
scales. This effect cannot be modeled by the two-dimensional simulations (see \emph{top-left panel} of 
Fig.\,\ref{structure_function_bayes}), while it is reproduced perfectly by the three-dimensional simulations (see \emph{top panel} of 
Fig.\,\ref{B cavities}).
However, since we know that the tapering imposed by the gas density distribution on the RM structure function at large 
separations may limit the possibility to determine the maximum fluctuation scale of the magnetic field, 
we may ask which is our actual sensitivity on $\Lambda_{\rm max}$. 
 
An accurate analysis would require considering $\Lambda_{\rm max}$ as a free model parameter in the three-dimensional
simulations as well, but the computational burden would be heavy. Indeed, we decided to explore only two more different values of
 $\Lambda_{\rm max}$, namely, a value of  $\Lambda_{\rm max}=10$\,kpc, which is lower than the best-fit value, and 
a value of  $\Lambda_{\rm max}=164$\,kpc, which is the maximum allowed by our computational grid of $1024^3$ pixels.

We performed the Bayesian analysis of the RM structure function
with the same magnetic-field configuration as is described in \S\,\ref{2-dimensional simulations},
except for the maximum scale of fluctuations. The results are presented in Fig.\,\ref{B_appendix}
for  $\Lambda_{\rm max}$=10\,kpc and $\Lambda_{\rm max}$=164\,kpc (\emph{left} 
and \emph{right panels}, respectively). In the \emph{bottom panels} of 
Fig.\,\ref{B_appendix} we present the two-dimensional
(colors and contours) and one-dimensional (histograms) marginalizations
of the posterior. 
The magnetic field central strength and radial decrease are constrained to
\begin{itemize}
\item $\langle B_{\rm 0}\rangle$=(33.1$\pm$9.7)\,$\mu${\rm G}, and 
  $\eta$=(1.7$\pm$0.4) for $\Lambda_{\rm max}$=10\,kpc;  
\item $\langle B_{\rm 0}\rangle$=(16.4$\pm$8.9)\,$\mu${\rm G}, 
  and $\eta$=(1.2$\pm$0.5) for $\Lambda_{\rm max}$=164\,kpc.
\end{itemize}
In the \emph{top panels} we show the observed
structure function (dots) along with the synthetic structure functions
from the posterior. The blue line is the best fit. 
We note that $\Lambda_{\rm max}=$10\,{\rm kpc} does not give a good 
description of the data because there is not enough power on large scales.

On the other hand, a maximum scale of fluctuation of $\Lambda_{\rm max}=$164\,{\rm kpc}
provides a better description of the observed structure function, even if it predicts
 slightly too much power for separations larger than $r_{\rm \perp}> 20$\,{\rm kpc}.
In this regime, the magnetic field model with $\Lambda_{\rm max}=$35\,{\rm kpc} is still
better.

Indeed, we may conclude that  $\Lambda_{\rm max}$ should be surely larger than 10\,{\rm kpc}.
Although the drop in the gas density limits our sensitivity, we may infer that $\Lambda_{\rm max}$
 should be around 35\,{\rm kpc}, possibly lower than 164\,{\rm kpc}.

\begin{figure*}[hb]
  \centering
  \includegraphics[width=17cm]{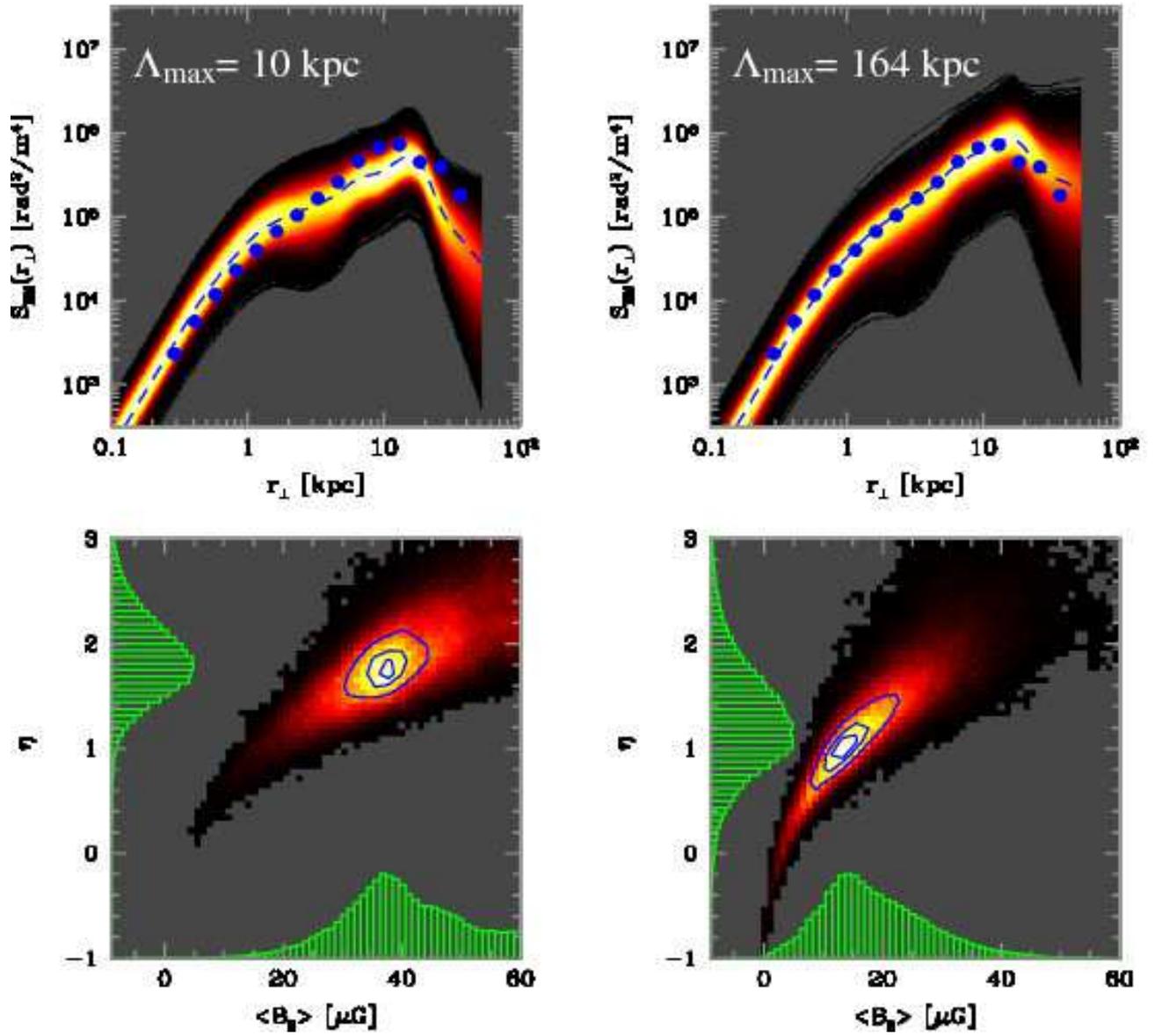}
  \caption{
Bayesian 3-dimensional analysis of the RM structure function for the 
  model with cavities for $\Lambda_{\rm max}$=10\,{\rm kpc} (left) and 
  $\Lambda_{\rm max}$=164\,{\rm kpc} (right).  \emph{Top panels}: 
  The dots represent the data (error bars are comparable to the size 
  of the symbols). The shaded area represents the population of  
  synthetic RM structure functions from the posterior distribution. 
  The dashed line corresponds to the most probable value for the model 
  parameters (see text). \emph{Bottom panels}:  one-dimensional (histograms) 
  and two-dimensional (colors and contours) marginalizations of the posterior 
  for the model parameters. The contours are traced at 0.9, 0.75, and 0.5 of the
  peak value.}
  \label{B_appendix}
\end{figure*}
\end{onecolumn}

\end{document}